# Estimating Primary Substation Boundaries and the Value of Mapping Great Britain's Electrical Network Infrastructure

Joseph Day, I. A. Grant Wilson, Daniel L. Donaldson, Edward Barbour, Bruno Cárdenas, Christopher R. Jones, Andrew J. Urquhart, Seamus D. Garvey

## Highlights

- A valuable open dataset has been created, a map of GB's primary substation areas.
- Aggregation of domestic energy consumption provides insight into decarbonisation.
- Suggestions are made to improve the consistency of geospatial electrical data.
- Extending this approach to all networks with monitoring enables digital twins.
- This allows for evaluating decarbonisation strategies at high spatial resolution.

## Abstract

Localised data aggregation in many countries including Great Britain (GB) is typically done to a geographical level with polygon boundaries that have a robust and trusted governance system in place. At a minimum this will mean there is confidence in a process to create a set of polygons that have unique identifiers coupled to geographical areas, and the ability to have these updated through a defined code of practice. Examples found across many countries are in the delivery of post, such as postcodes and zip codes, and of the definition of census areas and municipal boundaries. The confidence in these boundaries allows different data to be aggregated by third parties, which itself provides greater levels of data over comparable geographical areas to enhance wider analysis and decision making. As helpful as these polygons are for certain types of analyses, they are not however specifically defined for energy systems analysis. Here we combine publicly available datasets published from the six regional electricity Distribution Network Operators of GB to produce a new geospatial dataset with 4436 unique polygons defining the areas served by electrical primary substations. An example is also presented of the use of these polygons to link postcode level open government datasets on domestic energy consumption (2015-2020) from the Department of Energy Security and Net Zero (DESNZ). This results in another dataset with energy statistics aggregated to the geographical areas served by each primary substation across Great Britain. The significant value of the data we generate is demonstrated by evaluating future heat pump (HP) demands, considering up to 71% adoption by 2050. While preliminary, our estimates suggest that domestic HPs could increase annual demands by up to 51% on primary substations. However, there is significant variation in the increase, demonstrating that certain primary substations and regions should be prioritised for capacity expansion. Therefore, we believe there is a compelling argument for countries to set up processes to create and update polygons that have a meaningful relationship to energy systems. This would allow more accurate energy systems analysis to be performed, ultimately leading to an accelerated or potentially lower cost transition to a net-zero world.

# Graphical Abstract

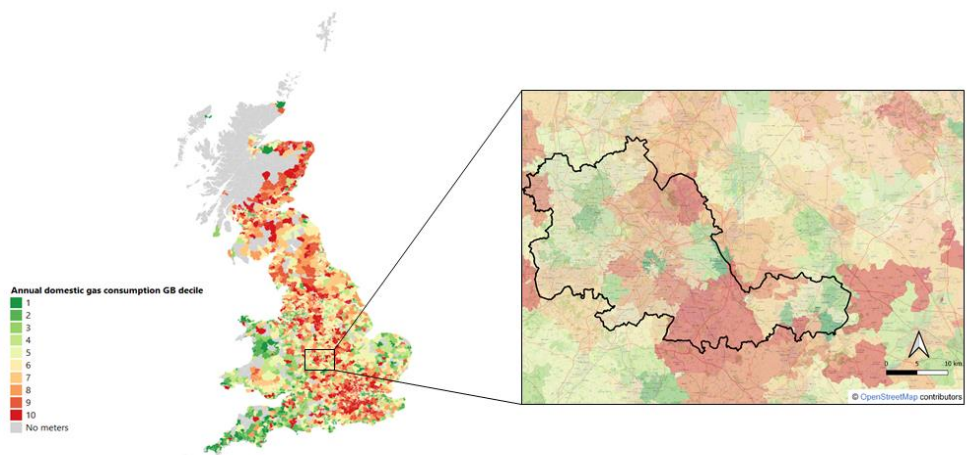

**Great Britain's domestic gas consumption mapped at a primary electrical substation level**

A single shapefile layer of all Great Britain's primary substations has been created for the first time. Geospatial and tabular domestic energy consumption statistics have been made available at this granularity for download at [44]. One use case of this dataset allows national and sub-regional variations in domestic gas consumption to be observed with reference to the electricity system geography as shown above for the West Midlands Combined Authority. These datasets (and others that can be created from these shapefiles) will be useful to inform the decarbonisation of heat in localised areas from a whole systems perspective.

# List of Abbreviations

COP – Coefficient of performance

DESNZ – Department for energy security and net zero

DFES – Distribution future energy scenarios

DNO – Distribution network operator

DOI – Digital object identifier

EPC – Energy performance certificate

EHV – Extra high voltage

ENWL – Electricity northwest limited

EV – Electric vehicle

GB – Great Britain

GWh – Gigawatt hour

HP – Heat pump

HV – High voltage

kWh – Kilowatt hour

LAEP – Local area energy planning

LSOA – Lower-level super output area

LV – Low voltage

MPAN – Meter point administration number

MPRN – Meter point reference number

NGED – National grid electricity distribution

NPG – Northern power grid

OA – Output area

OFGEM – Office for gas and electricity markets

OGP – Open geography portal

ONS – Office for national statistics

OS – Ordnance survey

QGIS – Quantum geographic information system

RIIO – Revenue = incentives + innovation + outputs

RUC – Rural urban classification

SPEN – Scottish power energy networks

SPF – Seasonal performance factor

SSEN – Scottish and southern electricity networks

UKPN – United Kingdom power networks

UPID – Unique primary identifier

UPRN – Unique property reference number

WMRESO – West Midlands Regional Energy System Operator

WPD – Western power distribution

# 1 Introduction
## 1.1 Introduction to the decarbonisation challenge

As the UK[1] aims to transition to a decarbonised energy system by 2050, it is widely accepted that data and digitalisation will play an essential role in achieving this goal across its wider energy systems [1-3]. There is an expectation of the growing use of near live to day-ahead data to help influence and manage the demand on energy networks, which provide heat, gas and electricity. For example, through measures such as time-of-use tariffs to influence demand and provide flexibility through the smart charging of electric vehicles (EVs) [4]. Additionally, geospatial data representing the locations of and areas served by pieces of energy infrastructure (e.g., an electricity substation) would be an important part of the data landscape for grid management as it enables more accurate spatial load forecasting [5]. There are also a range of helpful historical data that are created with a much larger window of analysis, e.g., annual gas and electricity consumption data, which enables researchers, local authorities and other stakeholders to evidence decisions for their local areas to decarbonise [6, 7]. However, these statistics are typically aggregated to geographical levels which are independent of the energy infrastructure including postcodes and census areas. Therefore, the predominant aim of this work is to address this knowledge gap by presenting a methodology which aggregates publicly available domestic energy consumption statistics to areas served by a common substation for the whole of GB. In doing so, this study demonstrates the value of substation areas as geographical units and their potential to aid in contrasting local decarbonisation options.

A suggested framework for helping the formation of feasible, sub-regional pathways to decarbonisation is through Local Area Energy Planning (LAEP) in which a combination of organisations (including gas and electricity network operators, local government and other experts in energy) consider the energy system in a more holistic way and produce a set of options, scenarios, priorities and actions that can be updated in light of changing knowledge and conditions [8]. A 2022 report from the Energy Systems Catapult stated that up to £252 billion could be saved from a co-ordinated and localised approach to the decarbonisation of heat, power and transport (as opposed to a centrally mandated, one size fits all solution) [9]. This kind of local area energy analysis is helped by better knowledge of the baseline demand from the existing energy system, so that proposed solutions and modelled effects are based on a starting point that has greater certainty.

## 1.2 Introduction to the decarbonisation of domestic heating

One of the single biggest sources of greenhouse gas emissions in GB is domestic heating at 14% of all emissions in 2020 [10], which is a focus of increasing policy attention [11]. A technology solution that aims to replace many natural gas fired boilers (which currently provide space heating and hot water for 85% of UK homes [12]) is through the deployment of electricity-powered HPs, described as 'the electrification of heat'. Notwithstanding the ongoing innovations to consider repurposing parts of the gas network to transport low carbon hydrogen and the potential zoning of urban areas for heat network development, it is a UK government policy aim for significant electrification of domestic heating through the deployment of household level air-source HPs or larger HPs connected to district heating. The UK government presently has a target to reach 600,000 HP installations a year by 2028 [13], a substantial increase from the 54,000 installed in 2021 [14].

---

[1] Northern Ireland, which is part of the UK but not GB, shares an integrated electricity system with the Republic of Ireland and is separate from GB's system. This paper focusses only on GB's electricity system, while using UK to refer to the central government and its policies.

Depending on the scale and speed at which the energy to provide heating is shifted from the natural gas system, there are expected impacts of both an increase in overall demand and of a potential change of timing of demand through the electrical infrastructure compared to historic norms [15]. The annual energy demand of natural gas for individual households with a gas connection and entire residential areas are typically much higher than annual electrical demand by a factor of 4 [16]. However, the COP of HPs means that only a fraction of the output for heat is needed as input electrical energy due to the amount of heat being harvested from the local environment. Despite this, there is an expectation that many existing electrical substation transformers and feeder cables, which make up the assets of the electrical network, might need to be upgraded to able to cope with this additional electrical demand. This is especially true when considering the extra loads from the electrification of transport (16.6% of new cars registered in the UK were EVs in 2022 [17]), and extreme cold weather events that reduces the diversity amongst the heating demand of consumers, leading to a higher peak demand over a day, as shown for the March 2018 cold weather driven spike in non-daily metered gas demand in Figure 1 that amounted to nearly 3.5 TWh. The equivalent electricity demand to provide this heat solely through HPs would likely be greater than the March 2018 peak daily electrical demand (i.e. > 1 TWh).

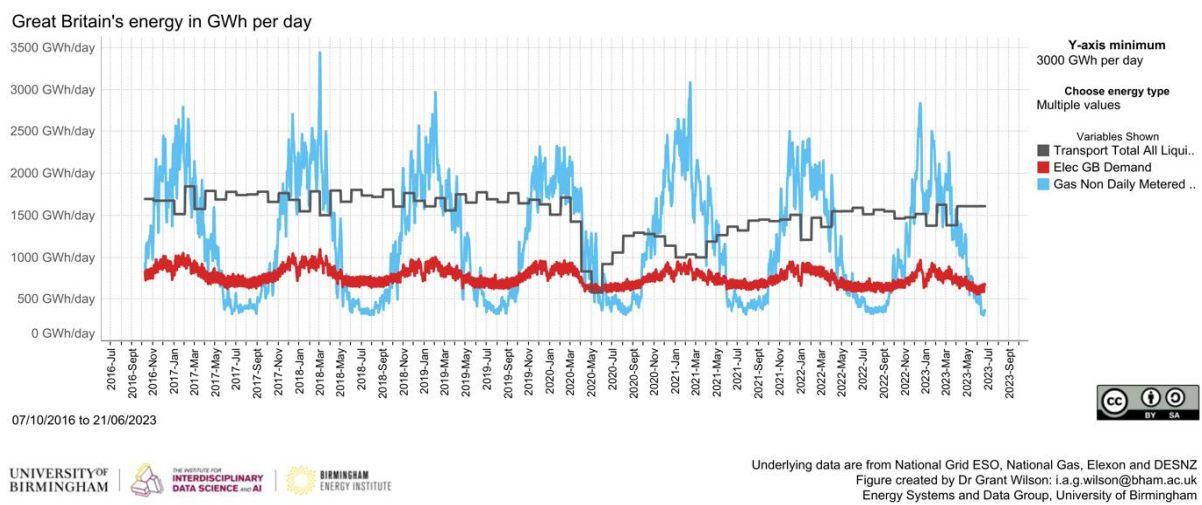

*Figure 1 – The multi-vector daily energy demand in GB from Oct 2016 – June 2023 with gas non-daily metered shown in blue, liquid fossil fuels for transport in grey and electricity in red. Switching major fractions of heating from natural gas to electricity would substantially increase the national electrical demand, particularly in cold weather periods. Data from [18].*

Therefore, a logical research question to be addressed by local energy planners would be, what is the domestic space heating and hot water demand over a period of time, presently provided by natural gas, in a given geographic area served by a distinct part of an electrical network such as a substation service area? The annual consumption would serve as a starting point to allow analysis to be conducted on the impacts on infrastructure serving a given area under different heat decarbonisation scenarios. By comparing the efficiency of gas boilers (typically 85% [19]) with the average COP over the heating season (known as SPF) of air-source HPs (typically 2-3 [20]), an estimate of the annual increase in domestic electricity consumption could be yielded based on the level of penetration of HPs under different scenarios. It would then require additional analysis to break down this annual value into a half-hourly time series but one that should be theoretically possible based on weather data and typical user profiles based on aggregated smart meter data or HP field trials [21]. Understanding the impact on the diversity of heat demand (both space heating and hot water) from extreme cold-weather events is an important challenge for designing a resilient, net-zero energy system [22].

The annual demand for domestic gas in a given geographical area would be broadly equal to the domestic gas space heating and hot water demand (since the only other residential use for the gas network in the UK is cooking, which represents less than 2% of annual demand [23]). To find the gas demand for an area, one must ideally know the annual consumption for each individual gas meter within the area. This information is held at the level of a unique meter point reference number (MPRN), by Xoserve (the UK gas sector's central data service provider) and available only to approved parties. Other secure methods of accessing this data exist but it is more open data that will likely have the most utility for LAEP creators due its ease of access. Fortunately, in this regard, small-area gas consumption statistics that cannot disclose individual data points (with sample sizes of at least 5) are published by DESNZ at the postcode level and are available for the public to download [24]. The annual domestic electricity consumption data landscape follows a similar format with household level data, for each unique meter point administration number (MPAN), maintained behind secure platforms (in this instance by Electralink, who are the data transfer service for the electricity sector) and postcode level data (for samples greater than 5) available for anyone to download from DESNZ [25] (in this case split into two datasets, one for economy 7 meters and one for standard meters). DESNZ also publishes annual domestic energy consumption statistics at higher geographies: census output areas, lower-layer super output areas (LSOAs), and local authorities[2]. These are useful as they can be combined with several other datasets which are released at LSOA level (such as data on fuel poverty [26] or socio-economic indicators from the census) or local authority level to give a quantitative comparison between areas represented by different elected councils. However, with regard to the preparation of a LAEP, an added value geographical unit would be one which is related to the electrical infrastructure itself, since it is the electrical system's assets that are going to be heavily impacted by the expected changes from the electrification of a large fraction of passenger transport along with the demand for space heating and hot water. We recognise that the geographical units relating to the electrical system as a less permanent structure than physical locations. Although the locations of substations are relatively permanent, their capacities can be upgraded and the feeder connections to customers can be reconfigured. Thus, the mapping process described here provides improved visibility of the requirements for upgrading and reconfiguration of the network.

### 1.3 Introduction to the electrical system in Great Britain

The electrical system in GB contains a transmission component (400 kV or 275 kV) and distribution level (all voltages 132 kV and below). Transmission level electrical infrastructure is operated by National Grid in England and Wales and in Scotland by two companies (Scottish Power Transmission and Scottish Hydro Electric Transmission) that are also distribution network operators. Transmission level networks typically transport electricity at higher voltages over longer distances across the country from major power producers to grid supply points where they connect to distribution networks. DNOs are responsible for distributing electricity through reductions in voltage until it connects to an individual electrical meter. Historically this transfer or 'flow' of electricity has been in a single direction, from larger centralised generation to end users; however in recent years it has been joined by an increasing number of smaller electrical generators (such as wind and solar farms) that are embedded directly in the distribution networks themselves. There are 14 electrical distribution network licence areas in GB, a legacy of the regional area boards introduced in 1947 under nationalisation. Today these are operated by 6 privately owned companies and their respective customer totals and regions served are shown in Table 1 and Figure 2.

---

[2] The census in England and Wales uses OAs (typically with 100-200 households) as a base geographical unit. These are subunits of LSOAs (usually 4-8 OAs per LSOA), which in turn are subdivisions of a local authority (an area of an elected local government). Scotland has a separate census which uses data zones instead of LSOAs.

| DNO licence number from Figure 2 | DNO licence area | DNO group | Number of domestic customers (millions) |
|---|---|---|---|
| 01 | Northern Scotland | SSEN | 0.7 |
| 02 | Southern Scotland | SPEN | 1.8 |
| 03 | Northern England | NPG | 1.4 |
| 04 | Yorkshire | NPG | 2.0 |
| 05 | North West England | ENWL | 2.1 |
| 06 | Merseyside and North Wales | SPEN | 1.3 |
| 07 | South Wales | WPD[3] | 1.0 |
| 08 | West Midlands | WPD | 2.1 |
| 09 | East Midlands | WPD | 2.3 |
| 10 | South West England | WPD | 1.3 |
| 11 | Southern England | SSEN | 2.7 |
| 12 | South East England | UKPN | 1.9 |
| 13 | London | UKPN | 1.9 |
| 14 | Eastern England | UKPN | 3.2 |
| **-** | **Total** | **All DNOs** | **25.7** |

*Table 1 – List of DNO licence areas by their DNO group (i.e. the company which owns them) and number of domestic customers (e.g. domestic electricity meters) served. Data from [25] and this work. The number in brackets in the first column denotes its area in Figure 2.*

---

[3] WPD was purchased by National Grid plc in the summer of 2022 and became National Grid Electricity Distribution (NGED) but was a separate entity during the timeframe covering the initial part of this research.

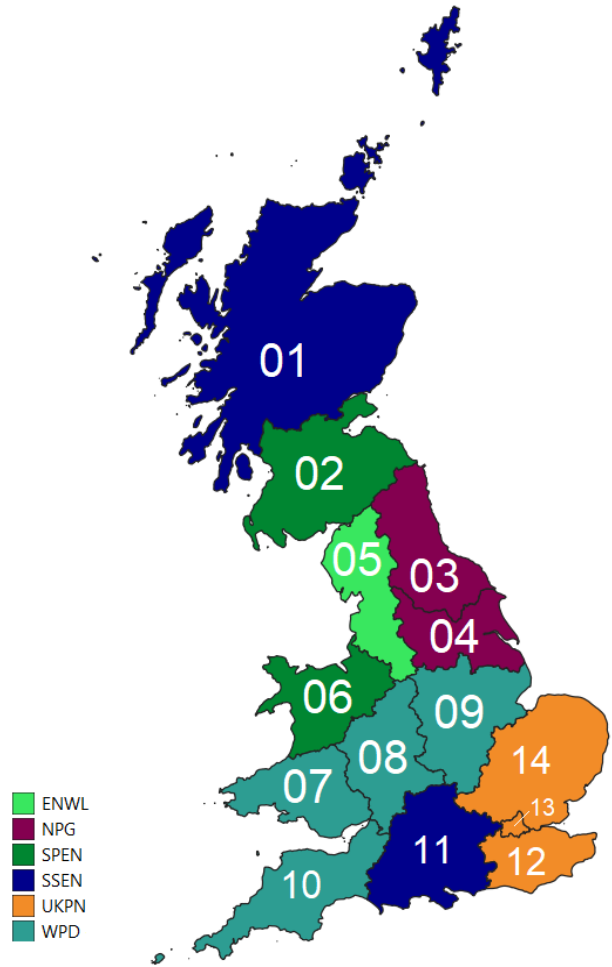

*Figure 2 – Map of GB showing the regions served by each DNO as per their number in Table 2. Data from [27].*

A typical hierarchy of an electricity distribution network is as follows and is summarised in Table 2. Grid supply points are substations which connect to the transmission level and reduce the voltage to 132 kV. Bulk supply points reduce the voltage further to 33 kV. Primary substations, also known as HV substations, then step down the voltage to 11 kV. Finally, secondary substations (also known as LV or distribution substations) reduce the voltage to 400 V (with 230 V per phase) which is suitable to be used in homes and small to medium commercial premises (some larger power users connect at higher voltages). Everything above primary substation level is sometimes referred to as EHV. There are exceptions as some areas of the UK use replacement levels of (e.g., 66 kV, 22 kV or 6.6 kV) due to legacy system designs.

Graph theory can be used to produce topological models of electricity distribution networks and can be a useful way to conceptualise them. In graph models, each node would represent substations, switches and the loads of end-users while each edge corresponds to a feeder that connects the nodes [28]. Depending on the voltage level, the topological configuration would vary. Within GB, HV circuits are typically connected in a complex ring network [29], although they generally operate radially with a break in the ring at the normal open point, and use two input feeders per LV substation to provide redundancy (an alternative path for power to flow in the event of a fault). On the other hand, LV networks mostly connect and operate in a radial fashion where each node is served by a single substation. LV substations typically have 1-6 output feeders in rural areas and 2-12 in urban settings [30]. This is the case because failures on an LV network affect fewer customers, so less resilience is required. Despite this, LV networks in some parts of GB are already operated in a meshed configuration, and increased LV network meshing (interconnections between circuits originating at distinct LV substations) has been proposed to accommodate greater penetration of low carbon technologies [31] and analysis of WPD's LV network showed end-users with the potential to be connected to another substation or LV feeder through normally open link boxes which can be closed during times of maintenance [32]. Nevertheless, irrespective of the voltage level and configuration, a geographical boundary can be drawn around the nodes representing end-users served by a common electrical infrastructure asset (such as a primary or LV substation) under normal network operation conditions.

| Substation | Input line to line voltage | Output line to line voltage |
| --- | --- | --- |
| Grid supply point | 400 kV or 275 kV | 132 kV |
| Bulk supply point | 132 kV | 33 kV |
| Primary (or HV) | 33 kV | 11 kV |
| Distribution (or secondary or LV) | 11 kV | 400V (230 V, line to neutral) |

*Table 2 – Summary of the voltage hierarchy and nomenclature of the GB electricity system.*

For maximum granularity and richness of data, the analysis of aggregating domestic gas and electricity consumption to a geographical area should be conducted at the smallest geographical level possible, which translates to as low a voltage level as possible. Ideally, this could be at the individual property (MPRN or MPAN) or LV feeder level. However due to data gaps, limitations of modelling or to protect privacy, it is likely these data will continue to be unavailable for LAEP analysis or the wider energy data community at this stage.

An enduring issue is that in 2023, it is still not possible for energy planners to access the precise shapefiles (geographical boundaries used in mapping software such as QGIS) of the smallest areas served by the lowest voltage electricity network assets, i.e., the service areas of LV substations. From the networks' own data, there is still some uncertainty regarding which LV substation or feeder serves specific customers [33]. Hence, in the absence of these datasets, the lowest voltage electrical asset with available shapefiles are primary substations which typically serve 5,000-10,000 domestic properties [34] with firm capacities (the maximum safe power flow they are rated to handle) of 2.5-30 MVA [35]. Another advantage of selecting primary substations is that as of 2023, this is also the lowest voltage level with good coverage of half-hourly power flow monitoring data; WPD have released this data under an open licence as of 2022 [36]. This means that present and future electrical flows can be considered under different decarbonisation of heating scenarios, using the available monitoring data from the DNO as a baseline. The forecasted power flows can then be used to determine whether network upgrades would be required at that voltage level.

There are several examples of where the geography of primary substations has been utilised for the benefit of the creation of a LAEP or to inform decarbonisation options. Distribution Future Energy Scenarios (DFES) use primary substations as areas of analysis to show projected uptake of low carbon technologies for each year until 2050 [37]. These aggregate statistics (and methodologies when made available) are useful for LAEP and other modelling but do not give a whole system view nor specifically say which assets will need upgrading and when. Within DNOs, data from primary substations is used to inform investment in the electrical network and the Long Term Development Statements and business plans which are part of the multiple year RIIO ('Revenues using Incentives to deliver Innovation and Outputs' [38]) frameworks defined by OFGEM (the UK energy sector regulator). These types of analyses conducted by the DNO are just for the electrical system, while a robust LAEP should include additional components of the energy network, including gas and heat networks for a more holistic systems approach. Examples of such projects are Net Zero South Wales [39] and the West Midlands Regional Energy System Operator project (WMRESO), the latter of which focused on the city of Coventry [40]. Said pieces of innovation work involved a collaborative consortium of partners from across the electricity and gas network operators, local government, third sector and universities producing plausible visions of detailed designs for future smart local energy systems in three scenarios. The three scenarios, much like the DFES framework, essentially differed in the extent to which hydrogen and electrification were deployed to meet energy demands presently provided by fossil fuels (high electrification, high hydrogen or a hybrid option). Both projects used primary electrical substations as a basic unit of geography for system designs, but this created issues when, as expected, the boundaries of other datasets such as local authority or census output areas did not align cleanly with electrical infrastructure. To assess the impacts of decarbonisation measures, datasets relating to energy (such as annual domestic gas consumption) and wider socio-economic issues (e.g., fuel poverty) are useful as baseline inputs and validation for modelling future scenarios. Well-formed data is needed to contrast the effects of various decarbonisation options on, for example, the economic impact on fuel poor households which is vital for a just transition [41]. The wider socio-economic datasets are often only openly available down to an LSOA level which does not overlap cleanly with primary substation boundaries as the two sets of areas follow a many-to-many relationship. As in WMRESO, an approximation can be made by taking the population-weighted centroids of LSOAs and assigning them to electrical primary substations on a many-to-one basis, which results in some loss of confidence in the derived primary substation attribute (e.g. fuel poverty rates within that primary substation area). These misaligned boundaries are demonstrated in Figure 3.

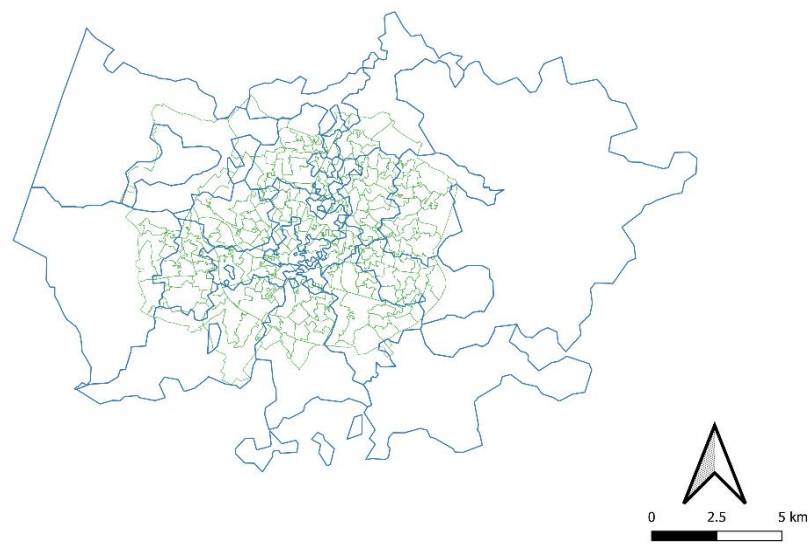

*Figure 3 – The primary substations that serve the city of Coventry, shown in blue, set against the LSOA boundaries in green. The LSOAs do not align cleanly with the primary substation areas. LSOAs outside of the city of Coventry are not shown.*

The same issue arises with annual domestic gas and electricity consumption statistics that are publicly available at the postcode level whose geographical areas can be assigned to primary substations on a much more accurately approximated many-to-one basis due to their smaller size. In some cases there is a naturally occurring similar geometry between postcodes and electrical network assets due to the LV feeders typically following the layout of road networks in urban areas. The Energy Systems Catapult have used this assumption to model the routing of both gas and electricity networks, although there are many exceptions, and a single road can sometimes be divided between many postcodes [42]. Tracing cables along roads has also been used by Barbour et al. to model community micro-grids [43].

### 1.4 Current study

Therefore, it becomes clear that the area served by a common asset of electrical infrastructure such as a primary substation would be a useful area for which to have insights and understanding. Although imperfect due to less LV network data availability, it is at present the best option. One important statistic regarding the energy system would be the annual electricity and gas demand across all domestic properties within a defined area. Hence, in the current study, a methodology was developed using the highest quality available public data to aggregate yearly domestic gas and electricity consumption at a postcode level to a primary substation area for all GB. The methodology also makes the data available for download under open access, along with the geospatial boundaries [44].

The aim of presenting and disseminating these data is to allow for the generalisation of this type of analysis (i.e., the aggregation of data to electrical infrastructure geographies) to include other statistics which are not presently considered by DNOs but may be of use to LAEP creators and other stakeholders seeking to design diverse decarbonisation plans for local areas. Also by sharing our methods, we hope to improve geospatial data standards and consistency across the DNO licence areas of GB. Our initial set of resources created this way could open up the process of scenario-based energy network modelling to a wider community and alert DESNZ or other national statistics collections (such as the OGP [45]) to the value of more data being made available at this spatial resolution.

## 2 Method

### 2.1 Collection of primary substation service area boundaries

The boundaries of primary substations were sourced from each DNO for their licence areas either through download via a public webpage or through a direct request to the DNO's data team. These took the form of shapefiles (files containing geospatial data) of file various extensions, so that analysis could be carried out in the open-source software QGIS and the geopandas library of python [46]. DNOs themselves are moving towards promoting open data practices and each have a platform where various datasets can be accessed and downloaded by external users. Table 3 presents the sources, file formats, availability, and methodologies used to create the primary substation service area boundaries across the six sets of files for the 14 license areas. Half of the files were available through public download at the time of the investigation in 2021-22 (although SPEN have since made their shapefiles publicly available online [47]) with the rest obtained by direct e-mail requests to relevant contacts within the organisation, who were identified by desk-research.

| DNO group | Availability | File extension | Split by licence area | Number of primary substations | Methodology to derive boundaries |
|---|---|---|---|---|---|
| SSEN | Via request | .shp | No | 861 | Voronoi (assumed) |
| SPEN | Public [47] | .shp | Yes | 722 | Voronoi (assumed) |
| NPG | Public [48] | .geojson | No | 582 | Output Areas |
| ENWL | Via request | .shp | N/A | 368 | Voronoi (assumed) |
| WPD | Public [49] | .gpkg | Yes | 1101 | Voronoi |
| UKPN | Public [50] | .shp | Yes | 802 | Postcodes |
| All | Public [44] | .geojson | Yes | 4436 | Mix of above |

*Table 3 – A summary of the shapefiles provided by each of the DNOs, the combined shapefile made by this work and the important properties of each file. Note that although ENWL, SPEN and SSEN did not explicitly provide methodology documents for the derivation of their boundaries, from the presence of straight lines, and comparing with WPD licence areas, they are very likely derived from a Voronoi polygons method.*

The DNOs had between 368 and 1101 primary substations within their area of responsibility. However, the methods used to derive geographical boundaries from the point co-ordinate locations of substations were different across the DNOs. The most commonly used procedure for mapping the service area supplied by a primary substation was through the creation of Voronoi polygons. These polygons are derived from a geometrical algorithm developed by Ukrainian mathematician Georgy Voronoy [51]. In this algorithm, a plane is divided into shapes based on the location of points such that for each point there is a polygon bounded by the region of space which is closest to that point. Put in alternative terms, the edges of each polygon will be the intersection of the lines of equidistance from two points. An example of a set of Voronoi polygons created from a range of points is given in Figure 4. Several DNOs use this procedure to create the primary substation boundaries using the point co-ordinates of the LV substations that are supplied from a primary substation to which they are normally connected. The first step is to create a layer of Voronoi polygons from the point locations of the LV substations and then, the polygons of all LV substations which share a common primary are merged into a polygon to give an estimation of the boundary for that primary. Using this method, it is possible to have a primary service area that has more than one polygon, e.g., when a primary substation service area contains an enclave of another primary this creates two polygons for the other primary substation that do not touch. WPD have validated these boundaries through a z-score analysis of the distances of LV substations from primaries and checking the outliers manually to verify the presence of errors in their network hierarchy. They state that "the methodology provides an accurate estimate of the geographic footprint of each distribution [LV] substation. After testing various input datasets, it was determined that determining Voronoi polygons at a distribution substation level provides a sufficient level of accuracy for the computational effort" [52]. WPD also used QGIS for their analysis and their boundaries were able to be partially recreated using the LV substations (whose co-ordinates can be downloaded from [53]) and the inbuilt 'Voronoi Polygons' algorithm which can be run in QGIS. However, because WPD did not provide the full LV to primary relationships at the time, the layer of primary boundaries could not be recreated in totality, but the component LV substation Voronoi polygons (which can be remade in QGIS from first principles) serve as a useful layer for visually comparing and verifying the primary substation service area boundaries. ENWL, SPEN and SSEN did not explicitly provide methodology documents for the derivation of their boundaries, but from observing the presence of straight lines, and comparing with the LV substation boundaries between WPD shapefiles and the ENWL licence area (who provided LV as well as primary substation shapefiles), it is highly likely that they were derived from a similar Voronoi polygons from LV substations method. The mapping of substations through Voronoi polygons has also appeared in academic literature [54-56].

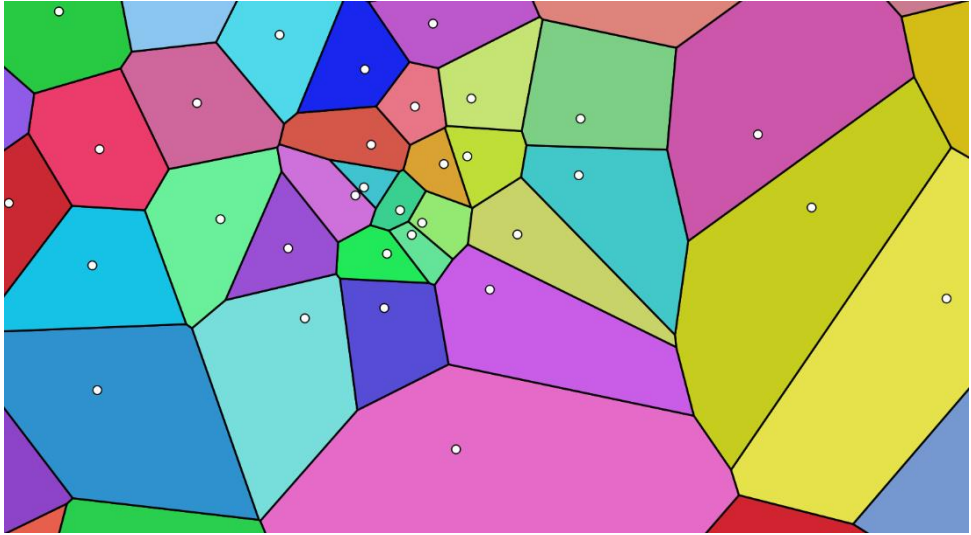

*Figure 4 – Example of a layer of Voronoi polygons where each edge represents a point of equidistance between two points. Taken from [52].*

The alternatives to using LV substation locations to derive service areas are LV feeder routes and customer addresses (provided the relationship from customer address to LV feeder is accurately known, which isn't always the case). UKPN have used the outline of postcodes to produce their shapefiles, from internally held information on the customer addresses and the configuration of their LV networks, i.e., which postcodes connect to which LV substations and the link between every LV and upstream primary (both in a many-to-one relationship).

In the case of NPG, shapefiles were drawn from a many-to-one relationship between census output areas (OAs) and primary substations. ODI Leeds (an external organisation) were commissioned to create the boundaries and used NPG-provided information on the relationship between postcodes and the electricity network layout, to allocate OAs to primaries on a many-to-one basis according to which primary had the most connections within an OA [57]. It appears, therefore, that NPG could have released the postcode-scale primary substation maps, but chose not to do this. During the analysis in this paper, both the UKPN and NPG shapefiles were verified to follow postcode and OA geographies by visually comparing their boundaries with OS's Code-Point-Poly [58] and OAs from the OGP [45]. The merits of using a purely geometric approach (such as the creation of Voronoi polygons) as opposed to a customer address-based method which uses existing physical geographies to derive geospatial files of primary substation boundaries are a matter for ongoing discussion. A major disadvantage, regardless of the method chosen by the DNOs themselves, is that there is no consistency or standardisation of approach. This is problematic when data is to be aggregated in GB, and to be able to be compared across different regions. It is hoped that our work will draw the attention of the sector to highlight this issue.

## 2.2 Combining the collected primary substation service area boundaries

In order to conduct analysis on the whole of GB at a primary substation level, the files compiled from each of the DNOs were loaded individually into QGIS. Some parsing of the data was required to fix the geometries of some layers as they initially returned errors when trying to run analyses. Four DNO groups (WPD, UKPN, ENWL and SPEN) had no errors in their geometries whereas SSEN and NPG had 3 and 52 errors respectively. The types of errors were as follows: too few points in geometry component, ring self-intersections and self-intersections. It was also required to modify the raw SSEN and NPG shapefiles as they did not differentiate between their two licence areas. For SSEN this was trivial as it could be clearly seen visually which licence area was Northern Scotland and which was Southern England. But for NPG, the two licence areas share a common border, so another set of shapefiles from National Grid [27] was used to badge each primary as either in the Yorkshire or Northern licence area.

Once these steps had been completed, all layers were combined into a single layer using the merge vector layers tool in QGIS to result in Figure 5 that contained 4,436 primary substation service areas. Each polygon in this new layer was given a unique numerical ID for each primary (referred to as the unique primary identifier or UPID). This is important from a robustness standpoint. The numbering convention to create the UPID was based on the DNO licence area number assigned in Table 1 and Figure 2, followed by a hyphen and 4 digits to allow for future additions and up to 9,999 primaries per DNO (for example, 01-0001 is a primary substation within the Northern Scotland licence area). Each DNO will have its own unique identifier for a substation, and it is hoped that overtime, there can be standardisation across GB for identifiers.

Unique identifiers are also vital to avoid ambiguity. If a string of text from the name of the primary were to be used, it would cause confusion as there were found to be 91 primaries with duplicate name strings. The files were also cleaned to remove unnecessary columns, leaving a shapefile with only the UPID, given primary substation name and DNO licence area. The geospatial file format chosen to be made available from the output of this work was the .geojson (standard RFC 7946 [59]) in the co-ordinate reference system World Geodetic System 1984 (WGS84).

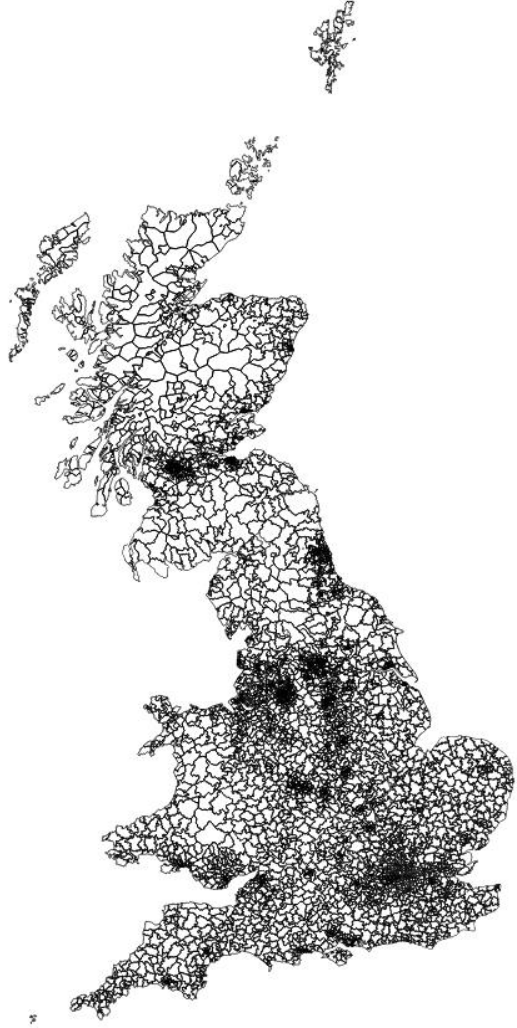

*Figure 5 – A map of GB's 4,436 primary substation service areas made by combining the datasets received by the six DNOs for all fourteen licence areas.*

## 2.3 Example use case of aggregating data to primary substation service area boundaries

In order to give a use-case example, the postcode level annual domestic gas and electricity consumption data were downloaded from DESNZ as a .csv file for each year from 2015-2020 [24, 25]. These were then linked to the primary substation service area geographies. The data included for each postcode was: the number of gas/standard electricity tariff/economy 7 electricity meters and mean/median/total annual domestic consumption of each meter type. The data comes from DESNZ who have access to the raw meter-level energy consumption data from Xoserve for gas [60] and from Electralink [61] for electricity. Further notes on the methodology of how these statistics are aggregated to postcode level and associated uncertainties are available at [24, 25].

The gas data was ready to use immediately but the electricity data required some modification. This was due to the electricity data having more than one file for each year, with postcodes split between standard domestic electricity meters and economy 7 domestic meters. Economy 7 meters give households access to cheaper electricity for 7 hours at night and were introduced in the 1970s to promote demand-side system balancing. As a result, the electricity consumption and electricity meter count for each postcode was summed across all types of meter, to give a total domestic annual electricity demand for each postcode (non-domestic demand is omitted from this analysis). When combined with the gas data, a single .csv file was now able to be used which contained all of the necessary information, with the gas and electricity consumption for each postcode from 2015-20.

In order to match and aggregate the postcode consumption data to a primary substation area, it was necessary to map these data so each postcode could be allocated to one (and only one) primary substation using the geographical boundaries received from the DNOs. To achieve this, the co-ordinates of the latest centroids for each postcode were obtained from the OGP with the National Statistical Postcode Lookup dataset [45]. These centroids could be loaded into QGIS and then joined on the postcode column with the parsed annual domestic energy consumption .csv (since the postcodes in each dataset were identical strings in the same UK postcode format). Once this joining of datasets was performed, the domestic annual energy consumption could be geographically located and visualised to points on a map. An example of this is shown in Figure 6 for the postcodes within and surrounding the city of Coventry). Then using this newly created layer of points, a spatial join operation was performed in python using 'geopandas.sjoin' from the geopandas library. This yielded a much faster processing time than the identical 'join attributes by location' feature in QGIS. The operation added a new attribute column to each postcode in the energy consumption .csv file, for the unique ID of the primary substation boundary in which it is located.

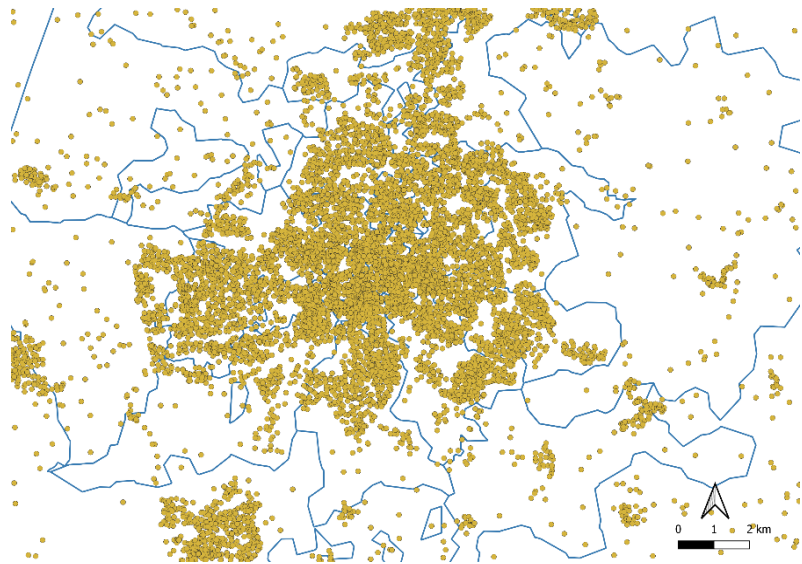

*Figure 6 – The postcode centroids shown as orange points against the primary substation polygons (with blue outlines) in Coventry. The use of postcode centroids allows for a many-to-one relationship between postcodes and primary substations.*

Once each postcode had been allocated to a primary substation, the totals of the consumption statistics given for postcodes (meter numbers and total annual domestic gas or electricity consumption) could be derived for each primary substation area by conducting a 'groupby' and 'sum' on the unique IDs of each primary substation column of the spatially joined dataset. This led to the creation of a new .csv file, where the index was the primary substation UPID, rather than the postcode, and contained useful columns such as the annual domestic gas and electricity consumption for each year 2015-20. The mean consumption by primary substation area was calculated (by dividing the total consumption by the total number of meters), as well as the means for each data item over the period 2015-20 (which are used as the values published in the dataset accompanying this publication and subsequent results section) and percentage changes from 2015-19 (accounting for 2020 potentially being an anomalous year due to the UK-wide lockdown beginning in March of that year). As the number of meters in the raw data can vary year-to-year, the mean numbers of meters for each primary were rounded to the nearest integer. Also, in a small number of cases (less than 25 out of 4436), the mean number of meters was less than 5 due to 0 meters being recorded at the same postcode for some of the years within 2015-20; in such instances, the number of meters across the period was set as 5 for consistency. In addition, an approximation of the percentage of off-gas properties per primary substation could be made by taking the ratio of gas to electricity meters (assuming that all domestic properties would have an electricity meter which should be valid for almost all of GB, since at most 0.3% of homes are estimated not to be connected to the public electricity grid [62]). In addition, the centroids of each primary substation polygon were calculated (using the 'Centroids' tool in QGIS) and considered against the local authority boundaries which can be downloaded from the OGP [45]. Replicating the earlier process of spatially joining postcodes to primaries using the postcode centroids (a point layer) to spatially join them to a polygon boundary, the primary substation centroids were allocated to a local authority. This was thought to add value to the dataset as local authorities in GB are administrative units under the control of elected councils to provide municipal services in their area. Therefore, by adding the local authority as an extra column for each primary substation, people who may use this dataset (which could include stakeholders connected with LAEP and local authorities) are quickly able to query it and list the primary substations in their area of interest.

## 2.4 Example use case of predicting future domestic electricity demand

Using these many-to-one primary substation-local authority area approximate lookups, for some future facing analysis, the National Grid's recently published 'Local Authority Level Spatial Heat Model Outputs (Future Energy Scenarios)' [63] (which contains projections of HP penetration at a local authority level in 5 year intervals from 2020-2050) was combined with this dataset and its present day consumption statistics to produce estimates of the changes in mean domestic electricity consumption by primary substation area. This was done by replacing part of the annual domestic gas consumption from the derived statistics with an additional electricity demand from the electrification of heat. Using gas as a proxy for heat demand (along with the ratio of the assumed boiler efficiency to SPF) determined the value for this additional electricity demand, which was then added to the existing annual domestic electricity consumption, to arrive at an equation for mean annual domestic electricity consumption in 2035 or 2050:

$$\text{Mean domestic electricity consumption } (future) = \\ \text{Mean domestic electricity comsumption } (today) + \\ \text{Mean domestic gas consumption } (today) * \text{percentage of properties on gas network} * \\ \text{percentage of properties with heat pump } (future) * \text{gas boiler efficiency} * \frac{1}{SPF}$$

*(Equation 1)*

In Equation 1, the percentage of properties with a HP in the year 2035 and 2050 for each local authority is taken from the Leading the Way scenario in [63] and the ratio of the relative efficiencies of a gas boiler to a HP is 0.315 (based on HPs assumed to have a SPF of 2.7 [20] and gas boilers having an efficiency of 85% [19]). An underlying assumption in this indicative model is that only existing gas boilers are replaced with HPs (i.e. it does not account for new build properties or replacing other fossil fuel-based or traditional electric heating systems with HPs).

Finally, population weighted LSOA centroids from the OGP [45] were used, firstly, to create an approximated many-to-one lookup from LSOA to primary substation area, then along with LSOA level

RUC [64], to determine the rural or urban nature of a primary substation area with regard to one of eight ONS designated classifications. If more than one classification of LSOA was present within a primary substation boundary, which was often the case, the classification was based on the one with the most occurrences.

To summarise, the methodologies detailed in this section were used to create a new geospatial dataset (available at [44]) which was augmented with additional publicly available data. The methodologies are summarised in the following process flow diagram (Figure 7). This allowed for the construction of the charts and maps in the following section.

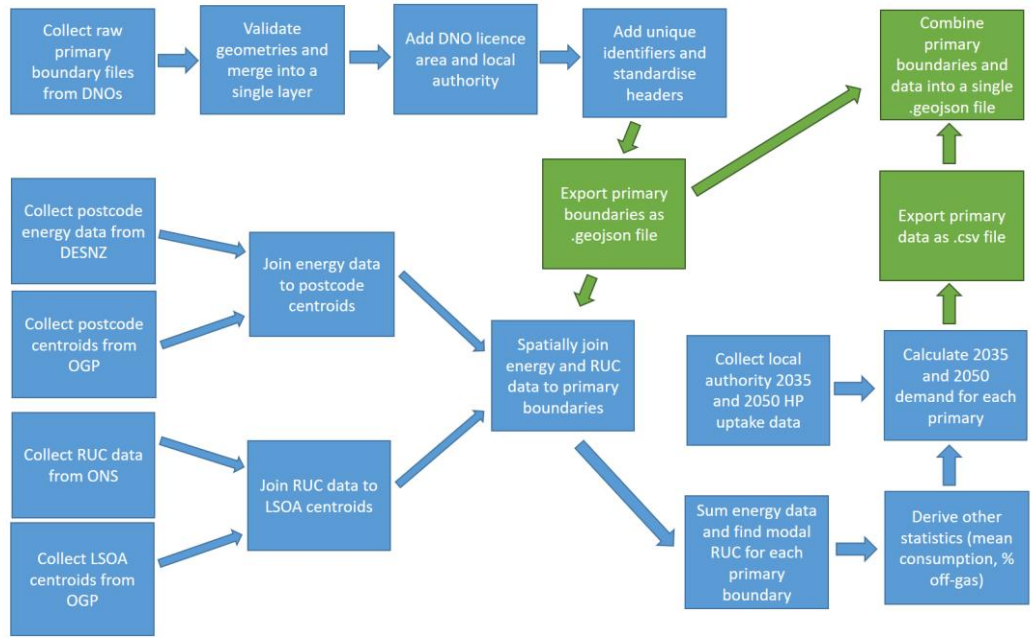

*Figure 7 – Process flow diagram to summarise the methods for creating the new dataset for GB's primary substation boundaries.*

# 3 Results

The raw datasets created through the method presented in section 2 can be accessed and downloaded from Zenodo in tabular and geospatial form [44] to allow the reader to query the data and produce their own visualisations of interest. Tables 4 and 5, and Figures 8-15 (consisting of maps and charts) convey some of the most relevant information from the datasets.

## 3.1 DNO licence area statistics

Tables 4 and 5 show a summary of the consumption and meter statistics for each DNO licence area and the total areas served by each of the six DNO companies. A typical licence area has of the order 1 million domestic electricity meters and a majority of its properties connected to the gas grid, although these characteristics vary across GB. Northern Scotland and South West England have the lowest proportion of homes with a gas connection (56% and 79%, respectively), while Yorkshire and North West England have the highest (both 95%).

The mean domestic gas and electricity consumption statistics also differ across GB; ranging for gas, from a minimum of 11,400 kWh in South West England to a maximum of 14,300 kWh in Northern Scotland, and for electricity, from a minimum of 3,290 kWh for Northern England to a maximum of 4,170 kWh for Northern Scotland. Domestic properties in most regions on average consume within 12,500 to 14,000 kWh of gas and 3,400 to 4,000 kWh of electricity, in line with GB mean statistics for the same time period (2015-20) [16]. These regional mean domestic energy consumptions are visualised in Figures 8 and 9, and although the reasons for these variations are multi-faceted and will be discussed in more detail section 4.1, the main factors which drive domestic energy demand are weather, income, total floor area, household size, connection to the gas network and the thermal performance of buildings.

| DNO licence number | DNO licence area | Domestic electricity meters (millions) | Domestic gas meters (millions) | % Domestic properties with gas meters | Mean domestic electricity consumption (kWh) | Mean domestic gas consumption (kWh) |
|---|---|---|---|---|---|---|
| 01 | Northern Scotland | 0.7 | 0.4 | 56% | 4170 | 14300 |
| 02 | Southern Scotland | 1.8 | 1.6 | 90% | 3430 | 13600 |
| 03 | Northern England | 1.4 | 1.3 | 93% | 3290 | 13700 |
| 04 | Yorkshire | 2.0 | 1.9 | 95% | 3430 | 13800 |
| 05 | North West England | 2.1 | 2.0 | 95% | 3590 | 13500 |
| 06 | Merseyside and North Wales | 1.3 | 1.2 | 90% | 3550 | 12600 |
| 07 | South Wales | 1.0 | 0.9 | 89% | 3390 | 12700 |
| 08 | West Midlands | 2.1 | 2.0 | 92% | 3760 | 13500 |
| 09 | East Midlands | 2.3 | 2.2 | 94% | 3750 | 13500 |
| 10 | South West England | 1.3 | 1.0 | 79% | 3880 | 11400 |
| 11 | Southern England | 2.7 | 2.3 | 86% | 3970 | 13400 |
| 12 | South East England | 1.9 | 1.8 | 94% | 3970 | 14000 |
| 13 | London | 1.9 | 1.7 | 84% | 3510 | 12700 |
| 14 | Eastern England | 3.2 | 2.8 | 88% | 4000 | 14000 |

*Table 4 – A summary of the domestic meter numbers and mean annual consumption per meter from 2015-20 for electricity and gas in each DNO licence area.*

| DNO company | Domestic electricity meters (millions) | Domestic gas meters (millions) | % Domestic properties with gas meters | Mean domestic electricity consumption (kWh) | Mean domestic gas consumption (kWh) |
|---|---|---|---|---|---|
| SSEN | 3.4 | 2.7 | 79% | 4010 | 13500 |
| SPEN | 3.2 | 2.8 | 88% | 3480 | 13200 |
| NPG | 3.4 | 3.2 | 94% | 3380 | 13800 |
| ENWL | 2.1 | 2.0 | 95% | 3590 | 13500 |
| WPD | 6.8 | 6.1 | 90% | 3730 | 13000 |
| UKPN | 7.1 | 6.3 | 89% | 3850 | 13700 |

*Table 5 – A summary of the domestic meter numbers and mean annual consumption per meter from 2015-20 for electricity and gas in the total areas served by each DNO company.*

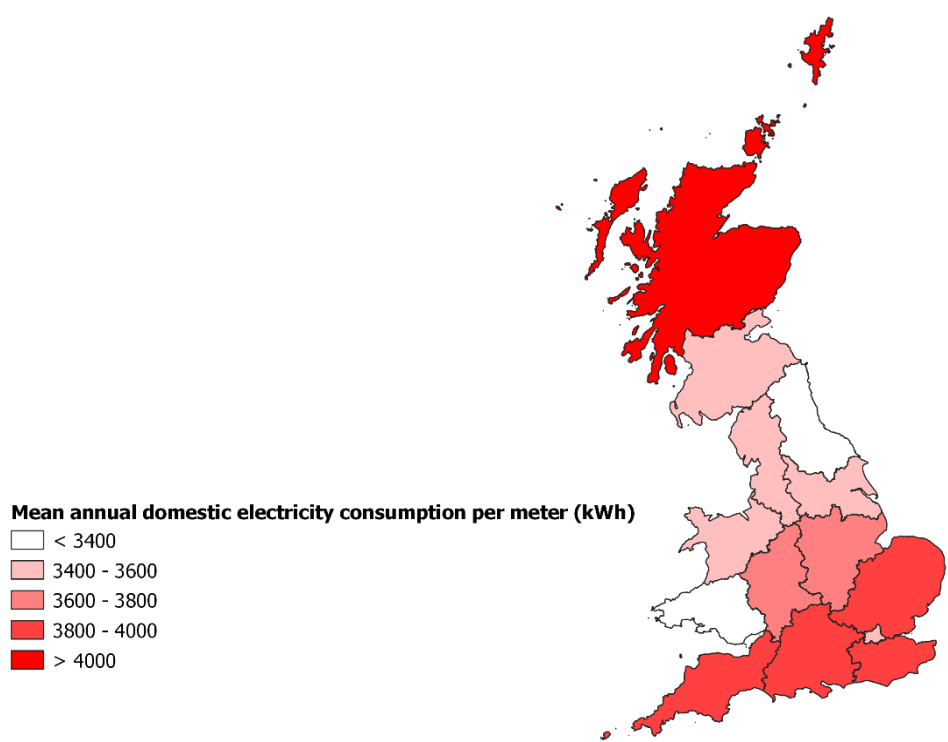

*Figure 8 – Map to show the mean annual domestic electricity consumption per meter (kWh) from 2015-20 for each DNO licence area.*

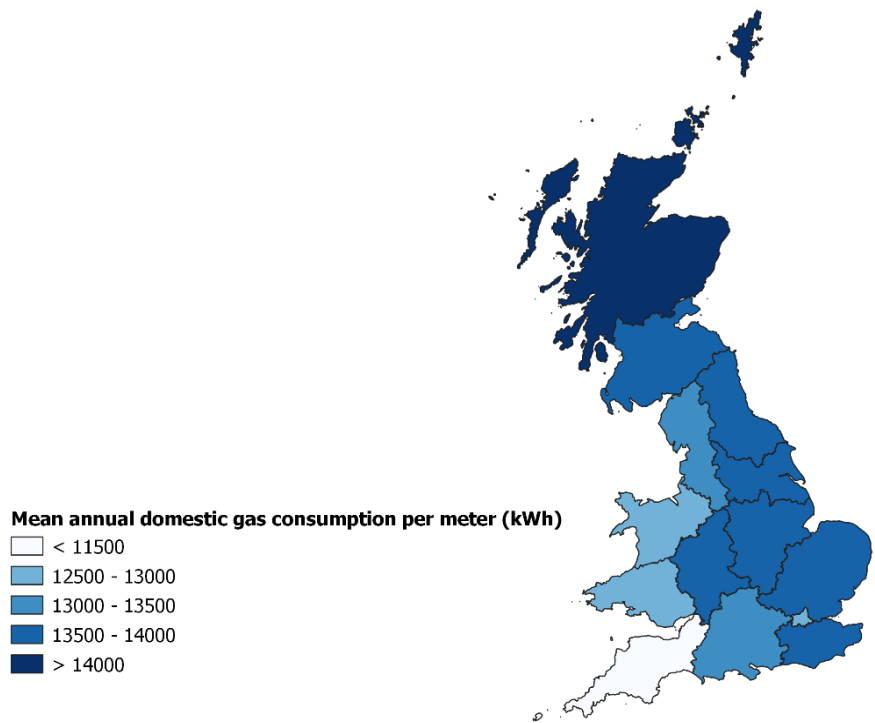

*Figure 9 - Map to show the mean annual domestic gas consumption per meter (kWh) for each DNO licence area from 2015-20.*

## 3.2 Primary substation area gas statistics

As well as comparing the domestic energy consumption statistics across DNO licence areas, it was important to aggregate them to a primary substation level in order to achieve the main aim of this analysis; to demonstrate the utility of the primary substation as a geographical unit. Once the dataset was created, visualisations could be produced to allow observations to be made at a granular level, meaning different cities and local authorities could be contrasted. One method to emphasise differences in mean annual domestic gas consumption per meter is to place each primary substation (with at least five consuming gas meters) in a decile from 1-10 based on its associated value (1 being the least consuming decile, and 10 the highest consuming decile). The English Indices of Multiple Deprivation uses a similar approach to rank and compare neighbourhoods [65]. A map for GB primary substation areas colour coded by their annual domestic gas consumption deciles is shown in Figure 10. Sub-regional variation can be seen (for example, the areas of continuous red in the London commuter belt). Similarly, the percentage of domestic properties not connected to the gas network in each primary substation area is shown in Figure 11, revealing a strong propensity for rural areas to be off the gas network[4]. More nuanced commentary on all these maps and statistics is provided in section 4.1.

---

[4] Appendix Figure A.1 maps the weather-corrected changes in mean domestic gas consumption from 2015-19 at a primary substation level, while Appendix Figures B.1 and B.2 give horizontal box plots for the distribution of total gas meters and mean annual domestic gas consumption per primary substation (for those with at least five consuming gas meters) for each DNO licence area.

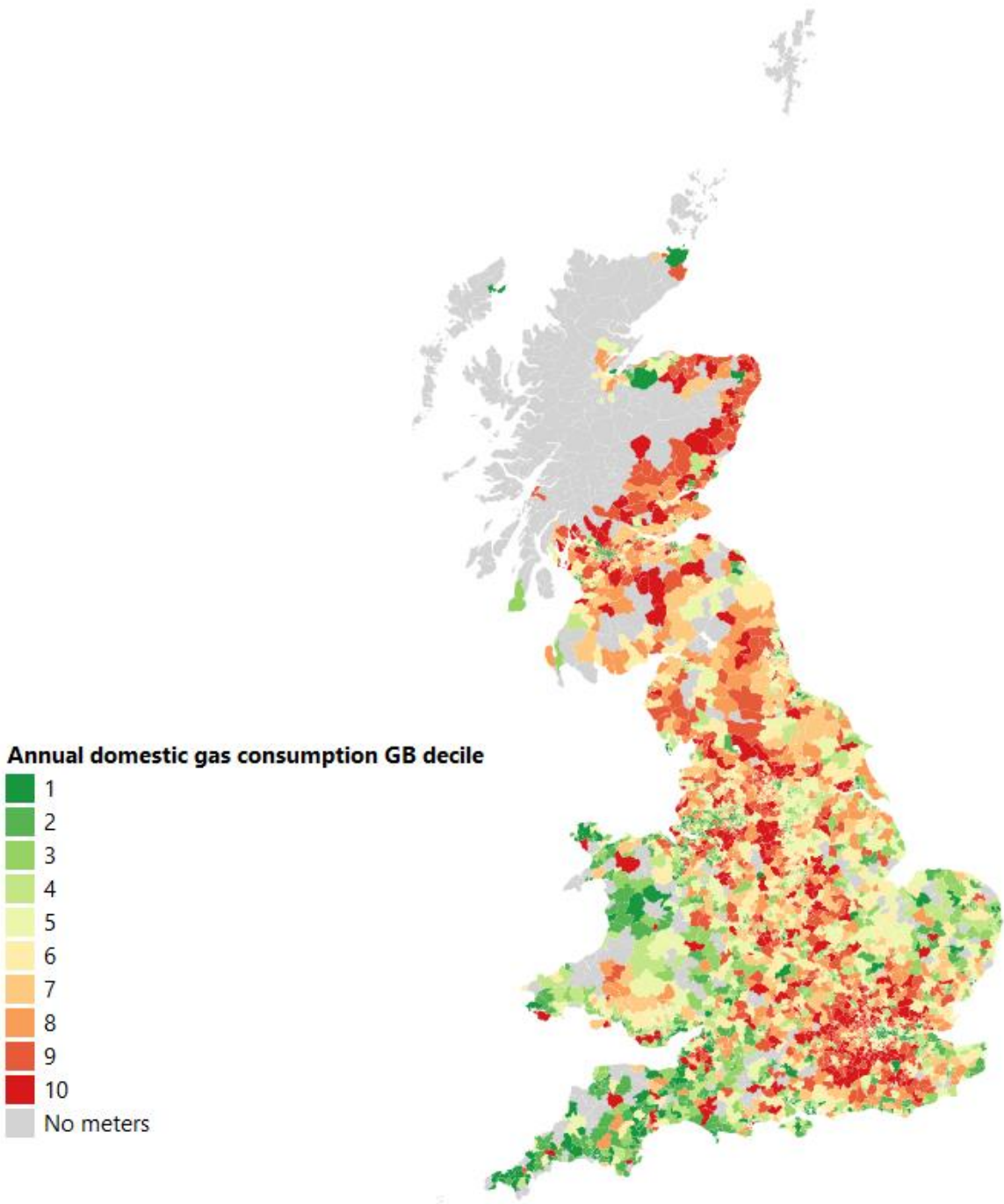

*Figure 10 – Map showing the GB primary substations by the decile of their mean domestic gas consumption per meter with the lowest mean consumption primaries in dark green and highest mean consumption primaries in dark red.*

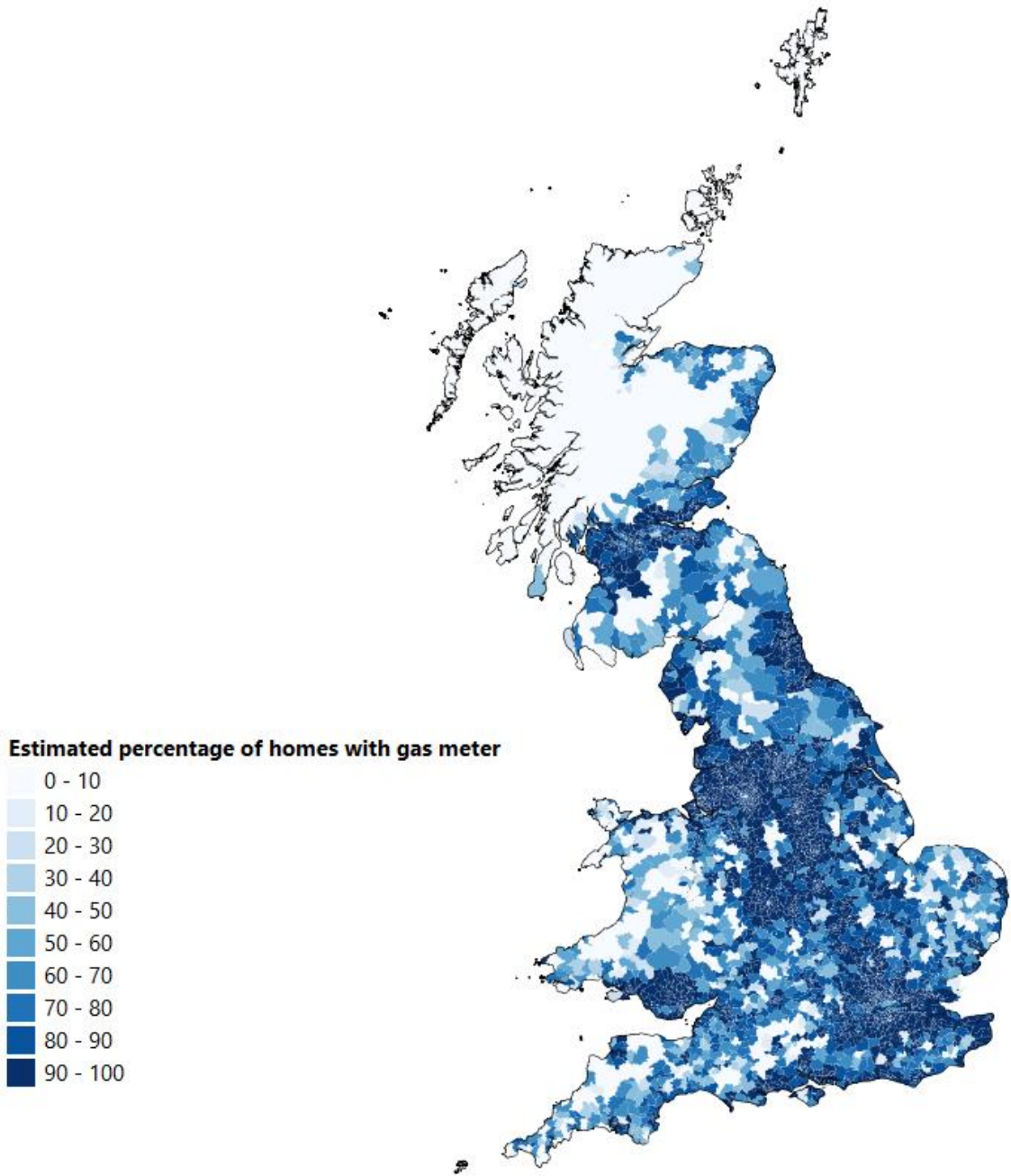

*Figure 11 – Map showing the estimated percentage of domestic properties connected to the gas network by primary substation area. The darker blue areas have a higher share of properties connected to the gas network.*

## 3.3 Primary substation area electricity statistics and projected future demands

In addition to domestic gas statistics, domestic electricity statistics are presented in the following horizontal box plots. These include the distribution of total electricity meters (Figure 12) and mean annual domestic electricity consumption per primary substation for each DNO licence area (Figure 13) for all primary substations with at least five consuming electricity meters. Concerning the likely future electrification of domestic heating in GB (outlined in section 1.1) and using the method described in section 2.4, the future annual domestic mean electricity consumption per primary substation area was estimated at 2035 and 2050, for HP penetration levels of up to 46% and 71% respectively. The distributions for the projected annual domestic mean electricity consumption per primary substation area are shown in the horizontal box plots of Figure 14 and Figure 15 respectively.

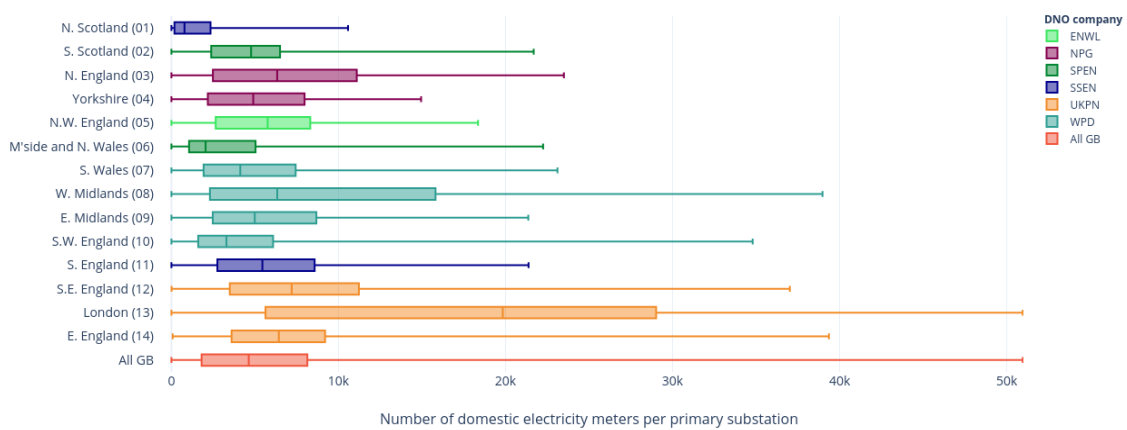

*Figure 12 – Horizontal box plot to show the distribution of the number of domestic electricity meters per primary substation by DNO licence area as well as for all of GB.*

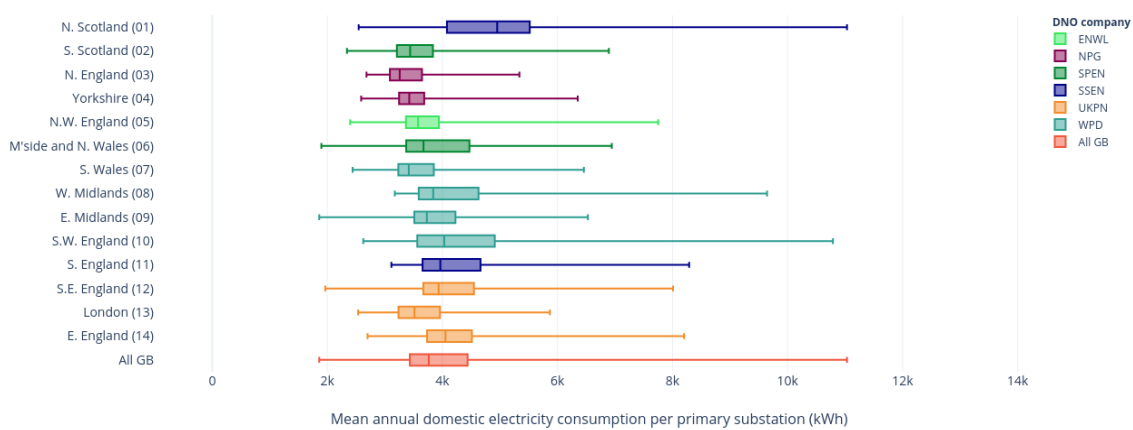

*Figure 13 – Horizontal box plot to show the distribution of the mean domestic electricity consumptions per primary substation by DNO licence area as well as for all of GB.*

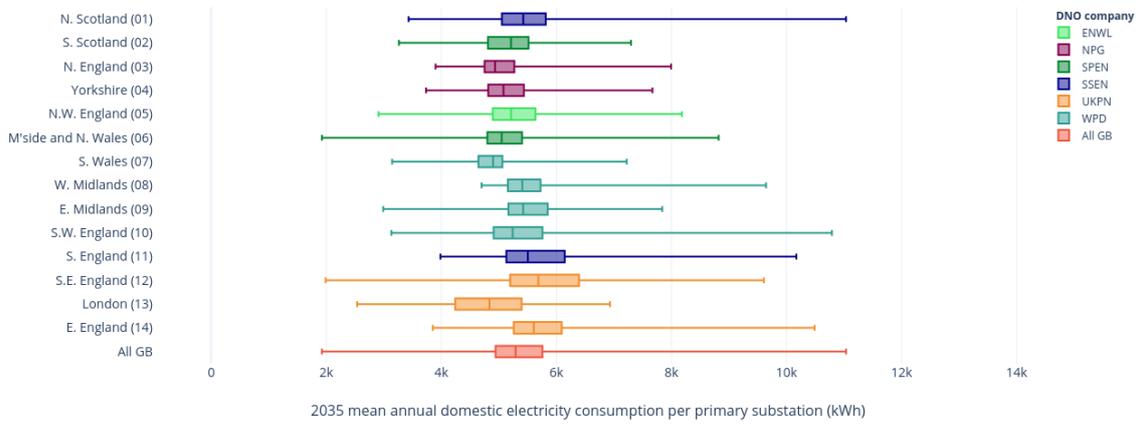

*Figure 14 – Horizontal box plot to show the distribution of the mean domestic electricity consumptions per primary substation by DNO licence area as well as for all of GB, under an electrified heating scenario in 2035.*

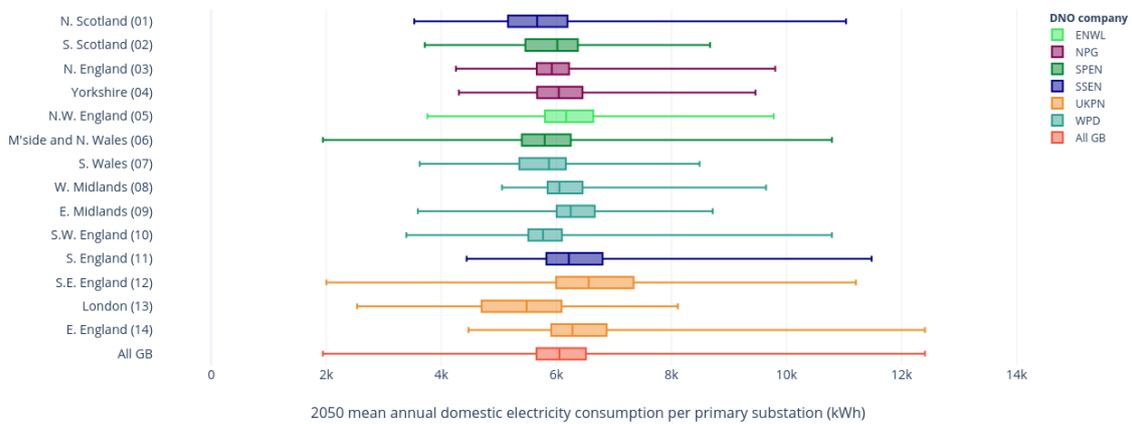

*Figure 15 – Horizontal box plot to show the distribution of the mean domestic electricity consumptions per primary substation by DNO licence area as well as for all of GB, under an electrified heating scenario in 2050.*

### 3.4 Rural urban classification of primary substations

The RUC of each primary substation area was mapped in Appendix Figure C.1, while domestic electricity meter totals per primary substation, mean primary substation areas (in km$^2$) and density of electricity meters (meters per km$^2$) for each of the 8 RUC are provided in Table 6. It is clear from the results that the primary substations in rural regions not only cover larger areas (which is fairly intuitive) but also have a much lower number of connected customers.

| Rural Urban Classification | Mean number of domestic electricity meters per primary substation | Mean primary substation area (km$^2$) | Domestic electricity meters per km$^2$ |
|---|---|---|---|
| Urban major conurbation | 10591 | 10.8 | 980.6 |
| Urban minor conurbation | 7067 | 12.3 | 574.6 |
| Urban city and town | 7764 | 31.9 | 243.4 |
| Urban city and town in a sparse setting | 4283 | 80.2 | 53.4 |
| Rural town and fringe | 3639 | 70.1 | 51.9 |
| Rural town and fringe in a sparse setting | 2630 | 149.0 | 17.7 |
| Rural village and dispersed | 1956 | 83.9 | 23.3 |
| Rural village and dispersed in a sparse setting | 1572 | 164.2 | 9.6 |

*Table 6 – A summary of the mean electricity meter numbers, geographical area and density of electricity meters for the subset of primary substations under each RUC.*

## 4 Discussion

### 4.1 Discussion: Results

The aim of this analysis was to show the value of aggregating energy statistics to areas served by a common electrical infrastructure asset. Firstly, when the mean domestic energy consumption data is aggregated to DNO licence areas, regional differences in domestic energy usage habits across GB can be observed. For gas, which is predominantly used in GB from October to April in the residential setting for space heating, it is unsurprising to note that the mean consumption per meter for each DNO licence area is heavily influenced by the climate of that region. GB experiences milder winters in the South and West of the country as shown in Figure 16 for 2017-19 of the mean winter temperature [66]. The combination of temperature with other factors (such as wind-speed and the previous day's temperature) strongly predicts gas demand at daily timescales and is used for the purposes of near-term system planning by the GB gas sector operators through a composite weather variable [67]. Despite these complexities, temperature remains (or more precisely effective temperature, which accounts for the temperature of the preceding day) the main driver of domestic gas demand; daily and sub-daily demands profiles have been calculated by Watson et al. based on effective temperature [68]. Consumer habits also influence demand, highlighting the combined socio-technical challenge of the decarbonisation of heat [69].

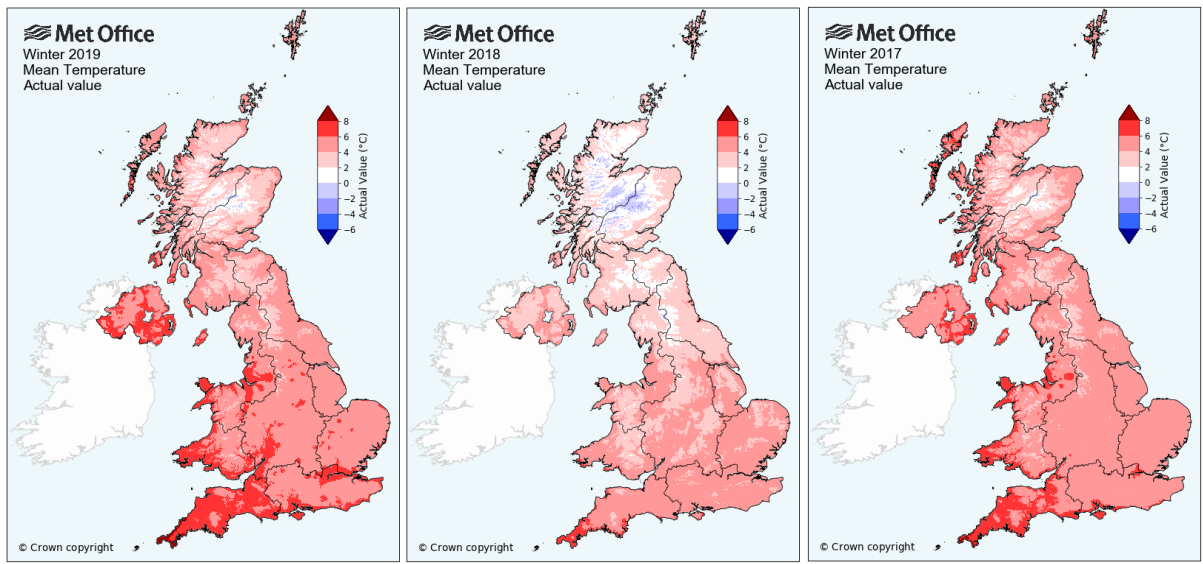

*Figure 16 – Map to show the winter climate's mean temperature of the UK from 2017 – 2019. Redder areas are milder on average, and thus can partially explain why mean domestic gas consumption is lower in these areas. Note the regional boundaries are not DNO boundaries but are the Met Office's 10 UK climate districts. Images sourced from [66].*

Therefore, Southwest England, which has large areas with regular mean winter temperatures above 6°C, is expected to have the lowest mean domestic gas consumption of 11,400 kWh, while Northern Scotland which often experiences winters with mean temperatures less than 2°C has a 25% greater mean domestic gas demand of 14,300 kWh. Moreover, a similar relationship (albeit less extreme) can be observed between Western and Eastern regions at the same latitude; e.g. South Wales vs. Eastern England and Merseyside and North Wales vs. the East Midlands. Within the regions, there will also be sub-regional variations based on microclimates and topography. However, it is important to note that climatic conditions are not the only variables driving domestic gas demand. There are other considerations such as: the types, floor areas and thermal performances of residential buildings; incomes, ages and heating behaviours of the people that inhabit the buildings and the efficiencies and available controls of heating systems. A report by DESNZ [70] has investigated these influences and derived a multivariate regressive model which stated floor area, building thermal performance and daily heating days as the variables of most relative importance (the latter of these will be strongly dependent on regional weather conditions). Nevertheless, many of the weather and non-weather factors affecting energy consumption will vary across GB by both DNO licence and primary substation areas. Investigating these relationships in depth is out of the scope of this paper. The aim is instead to highlight the importance of the aggregation or collection of data relevant to the energy sector at the geographical level of an area served by common electrical infrastructure asset such as a primary substation.

Domestic electricity consumption is partially influenced by heat demand (which in turn is affected by weather conditions) but less so than gas usage because electricity is much more varied in its end uses since it also powers lighting, appliances and home electronics, and increasingly electrical vehicles. While in the UK, 76% of domestic sector gas is used for space heating, only 17% of domestic electricity is consumed for this purpose [71]. Despite this, for houses with no gas connection and an electric source of main heating, the percentage of electricity used for space heating is much higher (estimated by DESNZ at 64% of the household electrical demand) [72]. Hence it is reasonable that for regions with a lower ratio of domestic gas meters to domestic electric meters, there would be a tendency for higher mean electricity consumption. This is the case in both South West England and Northern Scotland (the two regions with the highest incidence of off-gas properties) that respectively have on average the 1st and 5th highest domestic demand for electricity. However, there are clearly other factors involved, such as winter climate, which explains Northern Scotland's place as highest mean domestic electricity user and South West England's lower position than some regions with a greater rate of gas-connected properties. Furthermore, as for gas demand, relative affluence along with the size and condition of housing stock influence domestic electricity consumption. The study by DESNZ [70] also focussed on domestic electricity consumption and found the amount of high-energy consuming appliances owned, number of household inhabitants and floor area to be the three highest contributors to a higher electricity consumption. Other influences included the occupancy of the property throughout the week; the presence of electric technologies for space heating, hot water or cooking; thermal performance of the building and ownership of various energy intensive appliances. It seems logical to assume a strong correlation between affluence and larger property size as well as ownership of high-energy consuming appliances (e.g. large fridge-freezers, dishwashers, tumble dryers etc.) with a propensity for those on high incomes to use such devices regularly without fear of overspending on utility bills [69]. Therefore, it is not a mere coincidence that the three regions with the lowest gross disposable household income in 2019 (Northern England, Yorkshire and Wales) are also the regions with the lowest mean consumption of household electricity [73]. Notwithstanding this relationship, it is still possible for larger homes to be in fuel poverty if they have low income and high heat demand.

The number of domestic meters for each area in Tables 1 and 4 are a slight underestimate of the total number, since only domestic meters were considered and not those for non-domestic properties. DESNZ does publish the number of domestic and non-domestic meters at local authority level and across GB, the ratio of domestic to non-domestic electricity meters is typically 10:1, so an order of magnitude estimate of the difference can be made. The lack of geographically granular non-domestic electricity consumption statistics have also been identified by the WMRESO project as a significant data gap [40] which presents a challenge for whole systems, local energy area planning.

The estimated percentage of off-gas properties for each primary substation area is shown in Figure 11; 681 primary substation areas (15.4% of the total) are found to be 100% connected to the gas network and 542 (12.2% of the total) had no gas meters, so are 100% off the gas network. From comparing with the other map for RUC (Appendix Figure C.1), it can be observed that (as is the case in GB and expected to be similar in those countries with a gas system), there is a strong relationship between how rural an area is and the percentage of properties that are not connected to the gas network. Large contiguous areas of Scotland, Wales, South West England and Eastern England are not connected to the gas network, with a patchwork of mostly off-gas primary substation areas throughout other parts of rural England. On the other hand, the central parts of large metropolitan areas also have a relatively higher proportion of properties off the gas grid as shown in Figure 17 for Greater London overlaid on Open Street Map [74] (the same observation is noted for several British cities). This is likely due to the central parts of urban areas containing much more flats (because of increased competition for land) which in turn are less likely to have a gas based central heating system; 40% of flats in GB do not have a gas connection compared to 15% of all domestic properties [75]. The map created from this methodology for estimated off-gas grid percentages of each primary substation area aligns well with a previous map created at local authority, LSOA and postcode level for DESNZ [76], as well as tabular local authority level statistics published by DESNZ [77].

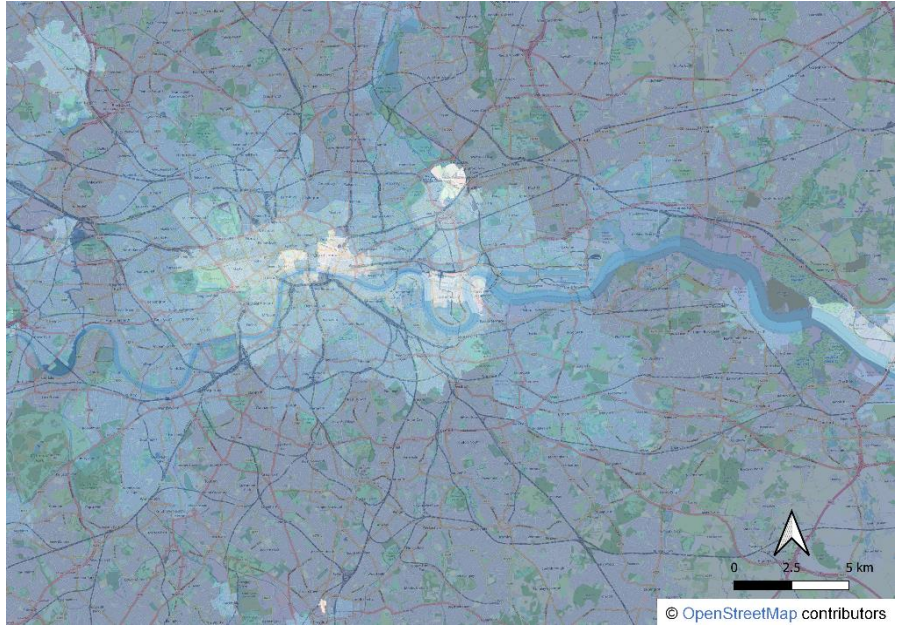

*Figure 17 – A map of the primary substations of Greater London using the same colour scheme as Figure 11 (lighter shaded blue areas have a lower rate of properties on the gas network and darker shaded blue areas have a higher rate of properties on the gas network). Inner London tends to have a greater proportion of off-gas properties than Outer London.*

It is observed from the horizontal box plots, Figures 12-15 in section 3.3 and Appendix Figures B.1 and B.2, that the UK's primary substations contain a diverse mix of attributes, in terms of geographical size, number of residential connected users and mean annual domestic energy consumption. These vary also across GB at the DNO licence area level. London has, on average, the greatest amount of domestic electricity and gas meters per primary substation, while Northern Scotland has by far the least. London also has the most extreme variation in the spread of meter numbers. By observing the shape and median of the 'All GB' plot at the bottom of each chart, some regions are shown to be more typical of GB as a whole than others. For example, for both the number of gas and electricity meters per primary substation, Southern England, East Midlands and Yorkshire most closely replicate the nationwide distribution as these regions contain a mix of rural and urban areas that is typical of GB as a whole.

From this analysis, the largest primary substation, both in terms of the count of domestic electricity meters and gas meters, was Fulham Palace Road C in the London Borough of Hammersmith and Fulham with almost 51,000 electric meters and 46,000 gas meters, covering a geographical area of 8.84 km$^2$. While by area (which can be calculated in QGIS for any polygon using the 'Add Geometry Attributes' tool), the largest substation area was found to be Grudie Bridge in the Highland local authority which covers 1,770 km$^2$ and the smallest Kerrera in Argyll and Bute with an area of just 87 m$^2$ (although as will be discussed below, this value is expected to be the result of an erroneous substation geography in the shapefile). Other primary areas, which are noted for potential inconsistency with the rest of the layer, are the 31 items in the NPG region named as '[Local authority name] direct to supply points'. These were very small areas (~1,000-10,000 m$^2$), often quadrilateral in shape, and it is implied that these are not in fact typical primary substations, but representations of areas connected to higher voltages (33kV and above) of the network (most likely large power consuming industrial sites). Similar small area polygons were found across the DNO licence areas for other large power users connected directly to 11 kV and with their own dedicated primary substation. Sometimes these can serve a surrounding residential area but often they will solely serve a single site.

In the latter case, a potential source of uncertainty will arise when any domestic consumers are included within its polygon and thus, it is difficult to determine the accuracy of the smaller substation boundaries and identify a smallest residential area primary substation. Nevertheless, these are edge cases and the mean GB primary substation was found in this analysis to serve an area of 51.1 km$^2$ that is strongly influenced by the urban or rural nature of the vicinity as shown in Table 6. Many primary substations were found to have no domestic gas or electricity meters; 130 had no domestic electricity meter (2.9% of primaries), 542 had no domestic gas meter (12.2%) and 119 had neither a domestic gas meter nor domestic electricity meter (2.6%). Some just had a handful of gas and electricity meters which could be due to the small area edge cases just described or the primary substation being in a very sparsely populated area.

The hypothesis that rural areas contain less customers per substation that urban areas is confirmed by Table 6. This shows that urban areas typically serve more properties on average than rural ones (almost an order of magnitude more for major conurbations compared to rural villages). The same trend across DNO licence areas was observed, while some uncertainty arose from the RUC data being unavailable for Scotland due to their devolved authority for these statistics and 217 primaries out the 3620 (6.0%) in England and Wales did not contain any LSOA centroids and therefore were unable to be assigned to an RUC (although many of these were very small, and as previously discussed, possibly erroneous small quadrilaterals representing some primaries).

Regarding consumption statistics, the largest total domestic electricity consumption was found to be within Verney Road in the London Borough of Southwark (total 163 GWh, mean 3,200 kWh) and for gas, Kingstanding in Walsall, West Midlands (588 GWh, mean 16,300 kWh). By considering the annual mean rather than total annual consumption, a more relevant statistic for determining the characteristics of a primary substation service area, the highest mean consumption for electricity and gas, were found in Abernethy Grid in Perth and Kinross at 11,000 kWh for electricity (n=14) and Peterlee Industrial in County Durham at 38,600 kWh for gas (n=5). However, these were based on very small sample sizes and possible allocation errors, so it is suggested to revise this analysis to only include substations with a minimum of 100 domestic meters. Under this minimum criteria the highest consuming substation by annual mean electricity consumption was Temple Farm in Windsor and Maidenhead with 8,300 kWh (n=402) and for gas was Warren Springs in North Hertfordshire with 28,900 kWh (n=139). Conversely, and applying the same filter on sample size, the lowest annual mean energy consuming primary substations for electricity and gas were CABLE ST T2 / CROWN COURTS T1 / GRADWELL ST T1 in Liverpool for electricity at 1,900 kWh (n=706) and BOW 11KV in the City of London for gas at 5,100 kWh (n=127). These extremes highlight the link between affluence and high electricity consumption as Windsor and Maidenhead has no LSOAs in the most deprived decile while Liverpool has the 2$^{nd}$ most LSOAs in the most deprived decile of any local authority in England [65].

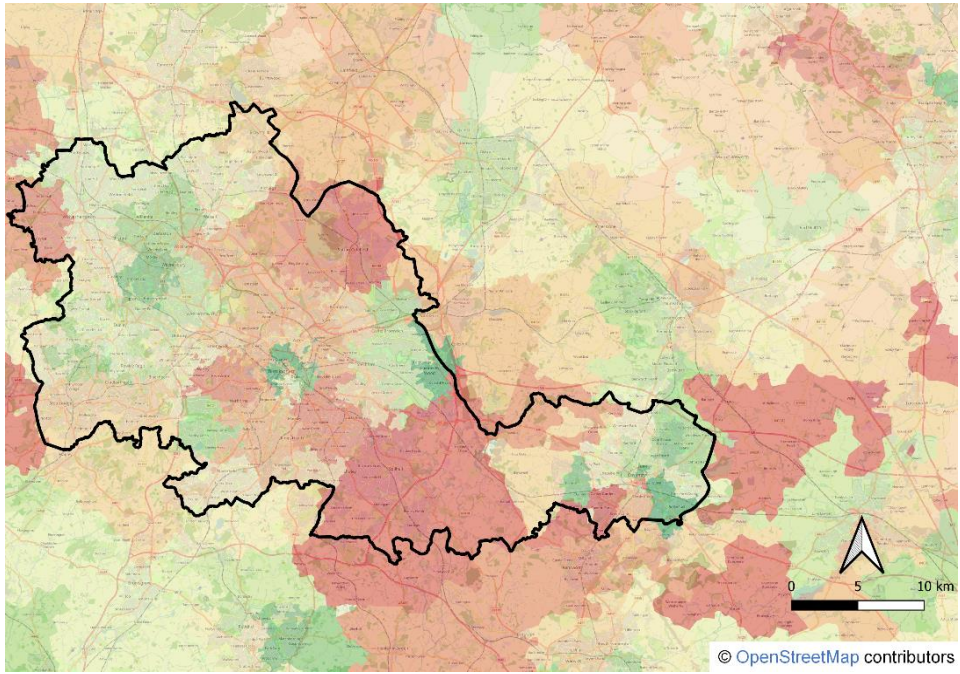

*Figure 18 – Map of the primary substation areas in the West Midlands Combined Authority (bounded by the black line) using the same colour scheme as Figure 10 to show the mean domestic gas consumption deciles. The red areas are the ones with the highest mean consumption and green areas have the lowest.*

In terms of the distributions of the annual mean gas and electricity consumption by primary substation, the horizontal box plots (Figures 13-15 and Appendix Figure B.2) provide further insight. The West Midlands, Southern England and South East England appear to most closely resemble the national picture for the distribution of the mean domestic electricity consumption per primary substation area, while for gas this occurs for Yorkshire, North West England and Southern England. An example use case is given in Figure 18 where a more transparent choropleth of the gas consumption deciles (as per Figure 10) can be overlaid against Open Street Map (in this case for the West Midlands Combined Authority and surrounding area), to allow navigators of the map to build their own understanding of an area, especially with regard to where greater challenges with electrical infrastructure may arise from the electrification of heat. This could eventually take the form of an online tool, similar to DFES maps published by the DNOs [78]. Moreover, localised maps of the trends shown in Appendix Figure A.1 could be produced to increase knowledge of the changes in consumption across a local authority.

From the datasets created in this study, primary substation areas on average saw their weather-corrected, domestic gas consumption increase by 0.9% between 2015 and 2019. This is well aligned with statistics published by the UK government which showed GB's mean weather-corrected domestic gas demand per property increasing by 1.1% over the same period, with year-to-year changes [79]. It is concerning that mean domestic gas demand has not fallen year-by-year in the late 2010s, which presents a barrier to the UK's climate change targets. This could be due to the slowdown in insulation measures installed and new homes being built with insufficient levels of thermal efficiency, an opinion echoed by the Climate Change Committee's 2022 report to the UK parliament [80]. Appendix Figure A.1 also appears to show a regional variation in the trend with eastern and northern parts of GB showing more frequent decreases. This could potentially be due to more insulation measures being installed in these colder areas which has reduced demand over the period 2015-19. Postcode level data is unavailable for the most recent years, although nationally weather-corrected domestic gas demand fell by 12% from 2022 as compared to 2021 [81]. This is likely due to the significant rise in gas prices and wider cost of living during that time period.

The horizontal box plots in Figures 14 and 15 show the distribution of the future mean domestic annual electricity demand per primary substation that would arise from the electrification of heat. At a GB level, the mean domestic electricity demand per primary substation increases from 4050 kWh today to 5390 kWh (+33%) in 2035 and 6130 kWh in 2050 (+51%). This rise is not uniform across the country as regions with greater domestic gas demand and higher instances of gas connected properties (such as South East England and Eastern England) experiencing larger relative increases. It is important to note, however, that these estimates are only indicative of one potential scenario. It is therefore likely that the real values would be even greater (especially in Northern Scotland and South West England) due to additional domestic electrical demand from EVs and HPs installed to non-gas connected properties.

## 4.2 Discussion: Uncertainties

Uncertainties and errors associated with the aggregation of data to primary substation level can be broken down into three categories: (i) geographic issues with the primary substation shapefiles themselves, (ii) limitations of the inputted DESNZ postcode consumption statistics and (iii) uncertainty around the future number of HPs installed and consumer practices.

### 4.2.1 Primary substation boundary uncertainties

There are different methodologies used to produce the boundaries for each DNO licence area (4 based on Vornois, 1 on OAs and 1 on postcodes). This creates inconsistency across GB regions and intrinsic uncertainty as the boundaries in all cases are only geometric estimates and not exact representations of the electrical network topography. An explicit example is shown in Figure 19 for the Voronoi derived boundary between two primary substations in the city of Coventry overlaid on Open Street Map, where the houses on Wainbody Avenue North are seen to abruptly fall between two primary substation service areas due the road being located close to the vertex of a polygon. In reality, it is highly unlikely that the houses on a continuous street built at similar time periods would be interrupted by such a change in electricity system topology.

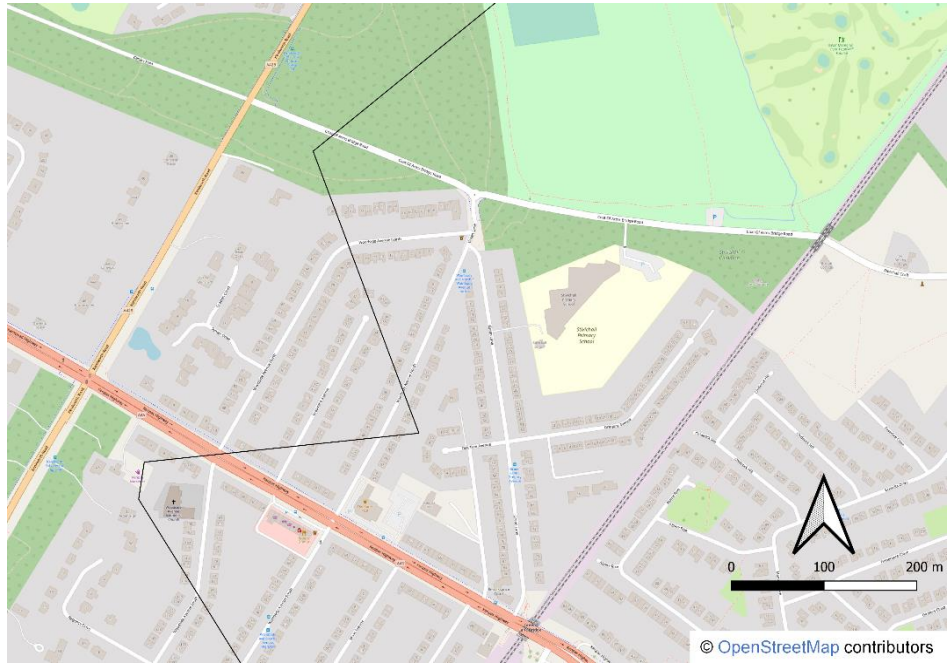

*Figure 19 – A boundary between two primary substations in Coventry derived through the Voronoi method. Note that the boundary cuts abruptly through Wainbody Avenue North which would be a highly unlikely electrical network topology.*

These issues, albeit perhaps less frequently, can also occur with OAs and postcodes because although they follow the road network (which in turn influences the electrical network itself in urban areas) in a more congruent manner, they still would have their discontinuities.

Within two of the DNO groups (SSEN and NPG) there are relatively small but occasional 'holes' within a single licence area served by the same DNO company. These could be a reflection of reality due to the area being very isolated (e.g. no electrical connections) or having its own electrical network supplied by an Independent DNO, or arise from a methodological error in the production of the shapefiles. An example of this is shown in Figure 20 for an approximately 25 km$^2$ hole within NPG's licence area. Some of these are in rural areas but others occur in urban residential and industrial zones. Kerrera in Argyll and Bute (Scotland), identified as the smallest area substation, was potentially erroneous, since it only encompassed a tiny strip off the Scottish coast and not the substation's namesake island. This could be solved (particularly in the special case of islands, although most are incorporated into the existing shapefiles reasonably well) by including all of the islands with a known grid connection as part of the input land mass for Voronoi analysis to be conducted on (with further manual Voronoi cleaning to occur if part of an island were placed into two substations that defied logical expectations). The concept of primary substation polygon cleaning could also be applied to other areas where one would not expect two different assets to serve the area. Digitalised datasets for the road network and other objects such as waterways, railway lines and private property land parcels can represent intuitive barriers for electrical cables and could be combined with the Voronoi algorithms (or other many-to-one geographical assignments, e.g. from postcode or OA to primary substation area) to supplement the assignment accuracy.

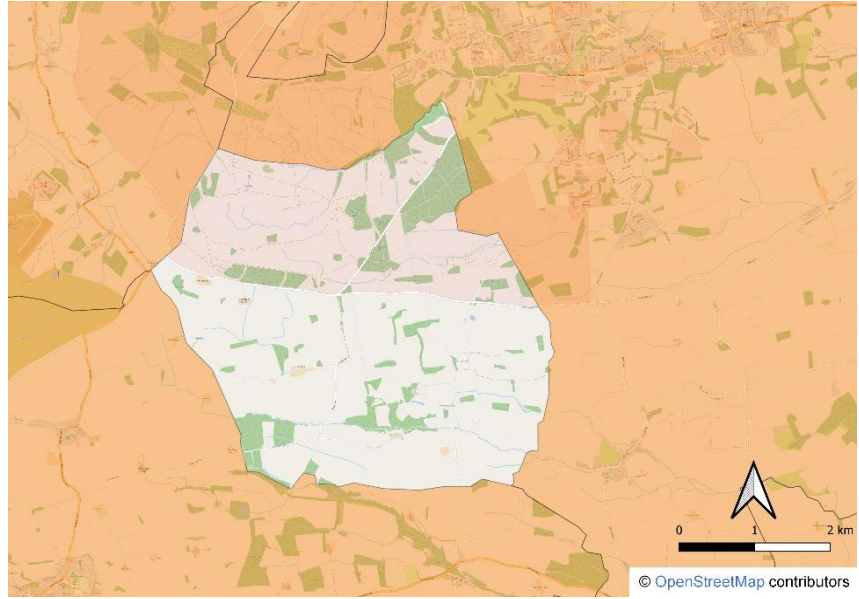

*Figure 20 – An example of an approximately 25 km² internal hole within NPG's primary substation polygon layer. The unshaded area has not been assigned to a primary substation area whereas the orange areas have been.*

Beyond the issues of geometry, it is also noteworthy that SPEN's Merseyside and North Wales licence area contains multiple site names and transformer abbreviations for its primary substation service areas (e.g. 'Primary Name 1' T1 / 'Primary Name 2' T2 / 'Primary Name 3' T3 as the name for one service area polygon, with T$n$ representing the nth transformer of a primary substation, and note that primary substations typically consist of a set of 2 or 3 transformers). This is because this licence area is unique in that it operates a much more of an interconnected meshed network at 11 kV, so transformers are linked to transformers from other primaries. This provides network resilience benefits such as reduced customer interruptions [82].

As well as geospatial polygon data for assets, other data structures for representing electrical infrastructure exist such as the Common Information Model, a data model promoted by OFGEM adhering to standards set by the International Electro-technical Commission [83]. Following queries in late 2022, GB's six DNOs networks are in different stages of progress with regards to publishing a Common Information Model. WPD have published one for public download on their website [84], while UKPN have produced another internal use only, citing data protection concerns (although they intend to release a reduced version for public use in the near future). On the other hand, the remaining DNOs viewed the challenge of creating a Common Information Model for all GB as one to be tackled through a working group consisting of OFGEM and all 6 DNOs. Hence, the sector's interest in creating data items consistent across GB supports the case for the importance of the foundational dataset created by this work.

In this study's dataset, geographical quality issues also arise at the boundaries between DNO groups and different licence areas within the same DNO group. At a boundary between two DNO groups, another hole may occur or the boundaries may overlap. An example of this phenomena is shown in Figure 21 for 3 substations in Dorset on the border between WPD's South West England and SSEN's Southern England licence areas. In this case, there are three instances in which the substation boundaries overlap (meaning a postcode centroid falling within this region would be allocated to two substations), and one area covered by neither of the substation boundaries (which means any postcode consumption data contained within this region would not have been picked up in this analysis). Furthermore, Figure 22 highlights two apparently distinct substation areas, one in the WPD West Midlands licence area and one in the WPD East Midlands licence area. However, on closer inspection, they are both called 'Coventry West', so it is unclear whether they are in fact the same substation or not, whilst it also demonstrates how (with the current data) the DNO licence area boundaries themselves (like municipal and census boundaries) do not cut cleanly across a same company's substations. For the purposes of aggregating demand from postcodes to primary substations, postcodes falling within no substation boundary were neglected and substations within two boundaries would be double counted within each of the overlaid primary substation service areas.

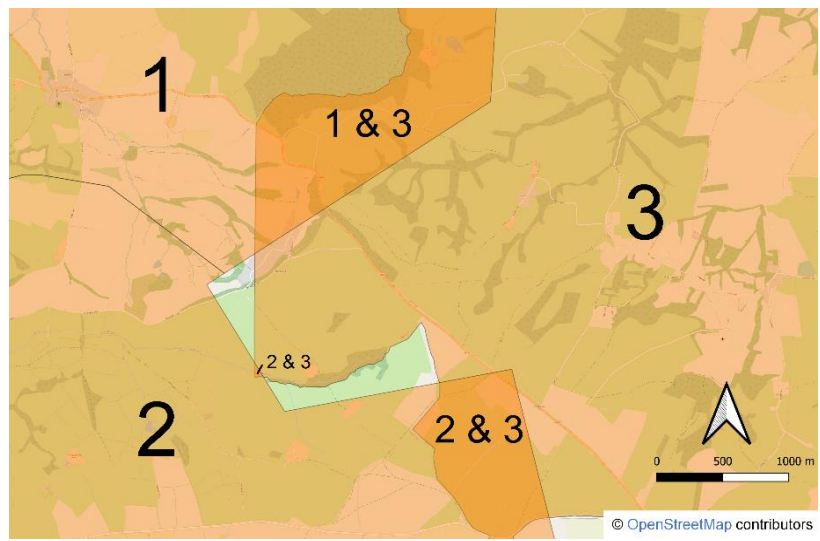

*Figure 21 – An example of an external hole in the created layer between the boundary of WPD's South West England licence area and SSEN's Southern England licence area. The numbers represent distinct primary substation polygons, so the areas denoted "1 & 2" and "1 & 3" are erroneously shown to be served by two primary substations.*

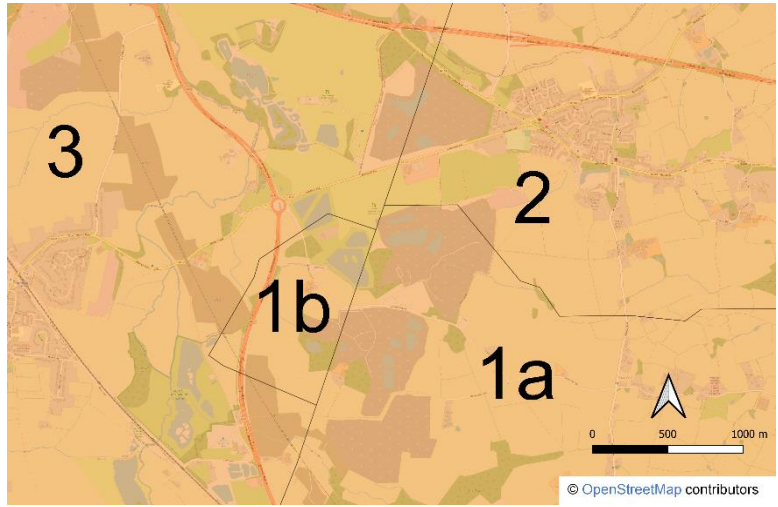

*Figure 22 – An example of a potential error between two licence areas operated by the same DNO (WPD West Midlands and WPD East Midlands). The numbers represent distinct primary substations, although 1a and 1b are both called "Coventry West" with 1a in the West Midlands and 1b in the East Midlands. It is possible they have been estimated to be served by the same substation but given separate polygons because they are in separate licence areas.*

All of these errors ultimately arise from the shapefiles being created through different methods by different organisations using incomplete data, or the fact they are made from geometric assumptions and not the actual layout of cables. One solution could be to combine all the DNO held data on the co-ordinates of all LV substations in GB (estimated to number 580,000 by the Centre for Sustainable Energy [85]) into a single point layer (along with their upstream LV to primary substation relationship), and then to recast the Voronoi polygon analysis on the whole of GB, which would eliminate the discontinuities shown in Figures 20-22. However, these data have not yet been obtained by our research group. On the other hand, if through greater data availability, the postcodes to primary substation relationships were known for the whole of GB, it might be more appropriate for this method to be used (along with OS's Code Point Poly) to create consistent shapefiles for the whole country. However, since these data might not exist for all DNOs and the postcode polygons are not yet available open data, the Voronoi method would appear to be a more sensible approach at this stage that would be uniform across DNO regions.

Another useful improvement would be for all networks to follow NPG's example of releasing geographical boundaries to represent the areas served by higher voltage levels (direct to supply point connections). This would provide a more holistic view of the network which considers customers connected at all voltage levels and their service areas across all of GB. In addition, this augmentation should be accompanied by manual cleaning to make sure that nearby lower voltage connected buildings are not included within the higher voltage substation boundaries. Finally, it is important to remember the dynamic nature of electrical networks. This means that the derived boundaries will not remain static over time, since varying degrees of non-radial network configurations allow for buildings to be connected to different primary substations at different times (e.g. during planned maintenance or potentially in the future through active management of parts of the electrical network). Also new developments will alter the network geography as the newly constructed buildings will require connections that can modify primary substation boundaries. Despite these issues, the dataset created by this work still demonstrates progress over existing public datasets which are not aggregated to geographies related to electrical assets and the network hierarchy.

### 4.2.2 Consumption data uncertainties

There are shortcomings in the analysed postcode level annual domestic energy consumption statistics published by DESNZ. Although for the purposes of this research, this is less important than the issues with the primary substation layer geometries. These secondary issues will be apparent in any imperfect dataset to be aggregated to a substation level, but it is worth exploring some of the issues specific to these results.

Firstly, the omission of statistics for postcodes with less than five domestic meters to prevent disclosure has a small effect on the accuracy of the results (although most consumption will be captured, since the overwhelming majority of households belong to postcodes with more than 5 domestic meters in total). This is less of an issue for gas meters, but more so for electricity where the splitting of the meter data into two classes (standard and economy 7 meters) creates a higher frequency of cases where one of the samples is less than five, so more of the data is missing. This leads to the intuitively illogical situation in parts of this analysis whereby the number of domestic gas meters within a primary substation was greater than the number of domestic electricity meters (one would assume almost every house in GB has an electrical meter, but this would not be expected for gas meters). This is the case in 637 of the 4436 primaries (although this tendency decreases in probability for the larger substations which will have more accurate boundaries and larger samples of meters).

Secondly, and perhaps related, due to the methods used to compile the DESNZ dataset, some meters are misclassified as domestic when they are in fact non-domestic and vice versa. For example, for gas meters a cut off of 73,200 kWh (approximately 5 times the typical annual household demand) is used to allocate meters consuming below that level as domestic and above that level as non-domestic. In reality, a very large domestic property or (more likely) a small commercial premises could be consuming an annual amount that puts them in the wrong category, leading to a potential source of error.

Thirdly, there arises a shortcoming in the methodology from using the postcode centroid to represent the entire postcode geographically, which was necessary to enable postcodes to easily map to primary substation areas on a many-to-one basis. This means some properties will naturally be assigned to the wrong substations, but it is a required sacrifice due to the limits of the substation boundaries available and the computational simplicity of avoiding having to work with polygons (whose 2-D areas might also lead to erroneous matchings if they are split over two or more primary substation areas). The best centroid to use for the postcode co-ordinates would be a population-weighted centroid, as has been created for census OAs by the OGP [45]. Thus, the creation of more granular, population-weighted centroids would be a useful resource for researchers.

Fourthly, there are some other intrinsic uncertainties associated with the consumption values themselves which can arise from: (i) the consumption values being based on estimated reads rather than actual measurements (this can be addressed with a greater penetration of smart meters with the functionality to automatically record consumption data at a half-hourly level, but as of 2023, still only 59% of domestic electricity meters and 54% of domestic gas meters operate in smart mode [86]); and (ii) a misalignment in the dates of measurement for the two energy vectors; the electricity consumption is over the period January to January each year, while the gas is over the period May to May. Also, importantly, the gas consumption data is weather-corrected, which means a correction factor has been applied accounting for weather conditions in that year, to allow more climate agnostic comparisons to be made between different years, whereas the electricity consumption is not weather-corrected. This means that changes in consumption driven by weather patterns will be more apparent in the electricity statistics, but have been suppressed by weather correction in the gas statistics. While this is not a major issue when electricity is not primarily driven by space heating demand, it will become more of a problem as heating is increasingly electrified through the greater use of HPs.

### 4.2.3 HP demand uncertainty

Finally, the HP analysis comes with many uncertainties by its very nature; i.e. a forecast might not always follow the actual situation that occurs. Although HPs are a mature technology with an accepted role to play in the decarbonisation of heat, UK policy remains uncertain [87] with large ranges in the estimates of hydrogen boilers and district heat networks in the future domestic heating mix. Across the three net-zero compliant scenarios in National Grid's 2022 FES [88], the percentage of homes predicted to be heated solely by a HP varies from 5%-34% in 2035 and 15%-52% in 2050. The 2021 UK Hydrogen Strategy [89] aims to reach a decision on natural gas pipeline conversion to hydrogen for domestic heating in 2026, following trials of a hydrogen village, while the 2020 Energy White Paper [90] set out a plan for implementing local authority heat network zoning by 2025 (whereby connections to a heat network could be mandated, if economically viable). Furthermore, the government has the "aspiration to upgrade as many homes as possible to Energy Performance Certificate (EPC) band C by 2035" [91], which would reduce the amount of energy required to provide thermal comfort in those dwellings. The consequences of these milestones will have a direct effect on the degree to which HPs increase domestic electricity consumption in 2035 and 2050. There are additional issues such as GB consumers' acceptability and acceptance of HPs due to their costs and perceived performance problems [92]. According to research by Nesta, only 25% of households asked would opt for a HP at their current price and the majority of respondents would not choose a HP over a gas boiler despite parity in price [93]. Moreover, the COP and SPF of different types of HP (e.g. whether it is air-sourced or ground-sourced) can vary significantly, especially during periods of cold weather. The effects of climate change might also impact demand for space heating in 2050 as well as cooling during the summer months [94]. Finally, annual domestic electricity consumption in 2035 and 2050 will ultimately depend on the rates of adoption of several other low carbon technologies including at-home electric vehicle charging and rooftop solar photovoltaics.

## 4.3 Discussion: Implications and Opportunities

Suggested future use cases and direction of this research are presented including: (i) the aggregation of many more datasets (public and private) to the geographical level of a primary substation, (ii) the extension of geospatial network data to include LV substation service areas (with exciting potential for detailed network modelling, and even potentially digital twins) and (iii) potential data governance structures to ensure the visibility, accountability and ongoing accuracy of this type of data.

### 4.3.1 Further datasets to be aggregated to primary substation areas

Despite the several described limitations of this approach due to the quality of data available, the process has been explained and could in principle be applied to any dataset with associated geographical co-ordinates. Future work could include analysis on EPCs at a primary substation level, which, since being made open data [95] and linked to Unique Property Reference Numbers (UPRNs), are able to be located by x-y co-ordinates (using OS's Open UPRN [96]). These certificates attempt to quantify the energy performance of a domestic or non-domestic building, producing an A-G rating based on the building fabric and heating, hot water, lighting, ventilation and cooling systems present. There are obligations to have a valid EPC in the UK for several legal reasons (such as when selling or renting a property) but their coverage remains at approximately 50% of all homes due to many properties never having required one. EPCs also contain methodological errors, due to incorrect entries by assessors and an assessment process that minimises cost based on old assumptions; one controversial outcome of which means a domestic EPC never recommends replacing a gas boiler with a HP [97]. Nevertheless, EPCs are still a useful source of rich data (the EPCs able to be downloaded publicly contain 90+ columns of information on topics such as the efficiency of building elements, presence of heating systems and tenure of a property), and thus can be a useful dataset to help gauge the ability of buildings within an area to retain heat. Therefore, compiling these statistics to primary substation service areas is a worthwhile exercise, since it allows network planners to know the thermal properties of buildings in an area, which is a significant driver for the annual energy demand for space heating. This could potentially lead to reduced network upgrade costs following increased electrical demand for heating, if a fabric first approach were to be taken. For example, in an area where HPs or an electric based low carbon district heat network are to be deployed, it could reduce the peak demand of electricity required to heat the homes and businesses, if their fabric is upgraded prior to or in parallel to any upgrades of electrical networks (which in certain scenarios might be a more cost effective option).

Furthermore, the UK civil service at DESNZ, and other departments (such as Department for Health and Social Care, Department for Work and Pensions and Department for Environment, Food and Rural Affairs), could also aggregate their extensive health, income and air quality datasets at the primary substation service area geography too. This would support a whole systems approach to energy planning that takes account of the interconnected socio-economic factors that drive energy demand in a geographic area as well as the parallel impact of energy injustice on health systems [98]. A practical outcome from these additional datasets being aggregated to electrical network boundaries could be additional knowledge around excessive fuel poverty or air pollution leading to investment in the electrification of transport or heating in a primary substation area. This could potentially help co-ordination between the local authority and DNOs, to consider where the trade-offs between network upgrade costs and public wellbeing benefit best align.

### 4.3.2 LV substation service areas and the art of the possible

Another area of future focus, where even deeper insights into energy systems could be gained, would be to continue to increase the geographical and temporal precision of the analysis. This would mean creating smaller geospatial boundaries to represent areas covered by lower voltage electricity network assets and also using sub-annual energy demand profiles. Exploring the smaller geographies first, the sub-primary level boundaries to be determined would be those for (in decreasing size) HV feeders, LV substations and LV feeders. Knowing the boundaries calculated after understanding the location of meters connected to the LV feeders, would allow for more accurate and robust network planning. However, the uncertainty surrounding which LV feeders that electrical meters are connected to is something that GB policy needs to focus on to allocate greater resources to answering, as the availability and quality of this data has been identified as a significant data challenge [99]. Many of the DNOs have made more granular geospatial network data available, either publicly or on request, such as the point location of LV substations or the geospatial location of the lines of HV feeders, although the area that these assets serve cannot be easily inferred. One crude approximation of the boundary of an LV substation is the Voronoi polygon (the same inputs from which a primary substation polygon is created). However, as the area served reduces, so does the level of certainty of the approximation; if the boundaries of each Voronoi are uncertain, this effect will become more apparent for an individual Voronoi rather than a set of Voronois for which only a small proportion influence the shape of the perimeter. WPD have made their Voronoi derived LV substation polygons publicly available [49] and ENWL provided them along with the request for their primary level boundaries. These can be useful for approximate analyses where many assumptions can be made, but to get the full picture, precise electrical supply area polygons must be used. Throughout this work, it was assumed that the data linking which properties connect to which part of the electrical at the lowest voltage levels must exist in some form for reasons of network safety and priority. For example, if there was a power outage or planned maintenance on part of the electrical network, the DNOs would need to know where their customers are located in relation to the network topology, especially for those households on the Priority Service Register who need reliable electrical connections, e.g. to keep medical equipment functioning. Also, there are heavy financial penalties for prolonged periods of disconnection that are levied by the regulator OFGEM [100], so again there was an expectation that DNOs would have detailed knowledge of which meters were connected to which parts of the electrical network. The reality, however, is that this data lies somewhere in the region of imperfect, and while it is good enough for meeting the historical needs of the DNOs, it does require increased focus to provide the most value for researchers, energy planners and the DNOs themselves.

WPD also offer access to a service called dataportal2 [35] after an approval process that provides a navigable and detailed interactive online map of their electricity system assets, in many cases right down to an individual property level of the electrical cable that connects to a meter. This map is incredibly detailed and very useful for locating cables (for example, before conducting digging works) and does allow a source of 'truth' to be found for the connections of individual properties or rows of properties by manually following the cables back to their respective substations. An explicit example of this is shown in Figure 23 for a neighbourhood in East Birmingham, highlighting the discrepancy between the Voronoi approximation and the manual drawing of boundaries using dataportal2.

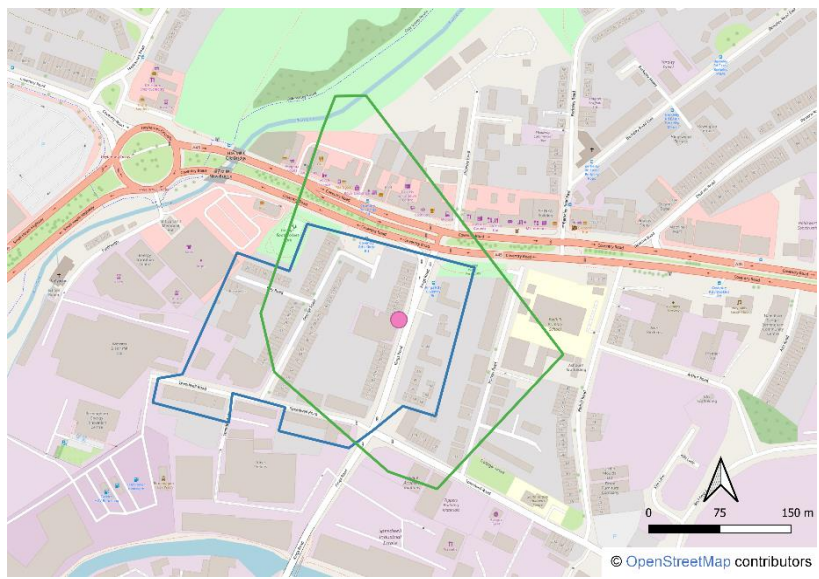

*Figure 23 – An example from East Birmingham of the differences at LV network level between the Voronoi method (green outline) of assigning substation boundaries and the actual truthful boundary (in blue) derived manually by inspection of 199 properties using an online interactive map [35] connected to the LV substation shown by the pink circle within the green and blue polygons.*

However, this manual method is clearly not scalable to do this for an entire region (at least not without a significantly resourced digitalisation team), and a bulk download option or ability to download the LV cables as a geospatial dataset does not unfortunately exist (only pdfs are able to be exported from the online interactive map). The challenge can therefore be viewed as one of digitalisation and openness; legacy datasets for network configurations which exist in a pseudo-digitalised form could be made available under licence conditions to researchers and those with an interest in local energy planning in the desired format. By working with the sector, necessary safeguards could also be put in place to mitigate concerns around privacy and security that come with making this detailed data more accessible, ultimately linking each building footprint or UPRN (or as many as feasible) to a corresponding LV feeder, LV substation and primary substation. Precise LV substation and feeder locations are essential for valuable ahead-of-demand network planning that goes beyond that which can be achieved from only a primary substation analysis, demonstrated by the insights that LV circuit upgrades have been predicted by Element Energy and Imperial College London to make up largest component of network reinforcement costs; 49%, 31% and 20% of these costs were attributed to LV, HV and EHV networks respectively by 2030 in their Core Decarbonisation scenario [101].

More granular time series data would also be useful for achieving peak demand analysis (a major influence for electrical system upgrades) because energy demands for GB households vary significantly across a seasonal, weekly and sub-daily basis. Annual level consumption statistics give one metric for which different areas can be compared but networks (of all energy vectors) are ultimately designed to handle instantaneous power flows (measured in kW), not the total amount of energy over a given time period (measured in kWh). This means it would be important to know the mean power demand flowing through an electrical system asset over a granular time period (such as every 5 minutes or half-hour). When considering future power flows, this type of data can serve as a baseline on which to add extra simulated demands from the electrification of heat and transport and to plan network upgrades or peak power flow mitigation strategies such as demand side response or various types of energy storage accordingly. The WMRESO project was given, at the time, privileged access to half-hourly transformer flow data for the 27 primary substations in Coventry. In combination with geographical boundaries for the primary substations, other datasets (e.g. on the typical floor areas of buildings) were able to be aggregated to that level and inform designs for future energy systems [102]. A similar framework is used for the DFES framework which in turn influences real network reinforcement strategies. Since then, this type of transformer flow data has been made more open for the WPD and NPG licence areas at primary level substations (the lowest voltage level with universal monitoring for WPD) and BSPs respectively [103-105]. However, even more in-depth models could be created if LV asset monitoring data was made available; as of early 2023, WPD have begun releasing sample LV substation monitoring data [106] and plan to monitor 10.9% of their LV substations by 2028 [107].

Open LV was an innovation project which sought to add community value by providing this kind of data for a given area of the network [108], whilst a wider rollout of LV monitoring features in the RIIO aligned business plans of many DNOs [109-113]. As a first step, the peak demand, peak generation, demand headroom, generation headroom, firm capacities and other attributes could be made available for all LV substations, even if they are estimates derived from internal DNO software. In conjunction with this, the release of new LV monitoring data (whenever it becomes available) and improved accuracy of LV substation and feeder service area polygons, could lead to the development of street level energy models and perhaps digital shadows (or eventually digital twins) by external organisations and the open modelling community, which could have an exciting impact on energy innovation. This would be enhanced further, if consistency across the DNO regions was promoted in their geographical allocation of properties to network assets and formats for time series data (such as ISO 8601 for timestamps to allow for ease of interoperability [114]). The ultimate geographically precise time series data would be that of the meter level for smart meters which record half-hourly consumption, but these are tightly controlled by the Data Communications Company for security and ethics reasons [115], so the raw, identifiable data is unlikely ever to be made available to researchers. However, anonymised data has been made accessible through a secure process (overseen by the Smart Energy Research Lab led by University College London [116]) and archetypes based on the mean from a large sample of consumers (e.g. a 3 bedroom, pre-1900 terraced house with a family of four) could be made available to researchers and would add value by allowing modelled demand to be contrasted with the measured demand from an LV substation (provided the building archetypes for each LV substation were known). The Centre for Net Zero's Faraday Tool [117] is another exciting development which uses sample data from customers of the energy supplier Octopus Energy to allow for the creation of archetypal demand profiles accounting for peak diversity, different dwelling types and the presence of low carbon technologies (including HPs, EVs and solar photovoltaics).

### 4.3.3 Data governance and visibility

Whilst considering the issue of data visibility and data governance, questions arise of where maps could be hosted, which licences could apply and who would be responsible for updating them on an ongoing basis. As an initial solution the platform chosen to host the data outputs of this paper is Zenodo as it provides a long-term repository for the data, creates a DOI for the data, and is able to provide other DOI numbers in future if an updated epoch of data is created. It therefore provides a snapshot of the primary substation service areas as they were when this research was carried out which can be referenced, and also has the functionality to accommodate updates in future. Each DNO has expressed their permission for the polygons to be shared under an open licence as long as each source was attributed. These data will therefore be publicly available and able to be downloaded by any interested party and under an open licence that means it can be used for any purpose as long as attribution is provided. Although the exact licencing arrangement of each component file used to create the combined dataset should ideally be specified at their original source as well; UKPN's primary substation files are explicitly mentioned to be shared under a Creative Commons licence (CC BY 4.0) [118]. Therefore, it is hoped that the sharing of this combined dataset under terms that align with CC BY 4.0 helps to set a precedent in the sector as a stepping stone towards more open data that does not have any licence conditions associated with it.

Going forward, a field of this importance merits agreement across the sector on approaches and who is responsible for its development. Not least because, recalling the operational nature of the networks themselves in which feeder circuits contain lots of interconnected nodes that can be turned on or off by switches, the boundaries of an electrical system asset are likely to become increasingly dynamic, so the shapes of the areas served would change over time.

A potential approach could be a platform akin to the OGP [45] maintained by the ONS or a similar organisation such as the Energy Networks Association [119] (the trade association to represent the common interests of all gas and electricity network operators) with support from a consortium of Universities. This platform would allow shapefiles of the areas served by different voltage level assets to be downloaded, and potentially different datasets to be queried and download for user-specified areas. Perhaps the functionality could include an interactive map that provided access to the datasets (with functionality to approve some that are user submitted) to be visualised at primary substation level. The Consumer Data Research Centre platform developed by the Universities of Leeds, Liverpool, Oxford and University College London, already produces and hosts several data-rich geospatial layers conveying a great deal of useful information about UK neighbourhoods in a user-friendly manner [120]. EDINA Digimap is also an exemplar of user-specified geographical areas for downloading data [121].

The responsibility and timeframes for updating the boundaries would need to be agreed across the DNOs to provide geospatial data on an ongoing basis. The regulator OFGEM could potentially help facilitate voluntary compliance with this agreement and encourage alignment with other ongoing projects such as the Energy Networks Association's National Energy Systems Map [119] or the National Underground Asset Register led by the Geospatial Commission [122].

As a contingency, if the data does not exist in sufficient quality and coverage within the data available to DNOs, an alternative could be to build on automated ways of assigning buildings, rather than specific electrical meters to substations. This has been tried by the Energy Systems Catapult's method of tracing buildings to substations along roads, as the electrical infrastructure is assumed to logically follow the path of roads in urban areas [42]. Although this method will have some associated errors, machine learning approaches based on sample areas could be used to iteratively improve the accuracy of these types of methods. Finally, although this paper has focused on both GB and electricity networks, it could also be extended into other countries with mature electricity networks. For example, Voronoi polygons have already been used to map service areas for critical infrastructure in the USA, Italy and China [54-56]. Also the concept of a geographical area being served by an energy infrastructure asset could be expanded to the gas system to facilitate a truly whole system approach to energy planning. In this case, a set of gas pipes which offtake the higher pressure pipelines at a common node could be thought of as an analogous gas supply area (similar to the service area of an electrical substation). This could have useful implications for aggregating heat demand and other energy related statistics to that geography, as well evaluating the impact of decarbonisation measures such as the impacts of decommissioning parts of the gas system in a given area or repurposing it to carry a low carbon gas such as hydrogen.

# 5 Conclusions

From the compilation of GB's electrical system's six DNO's primary substation service areas into a single digitalised map, the value of the primary substation service area has been demonstrated as an important geographical unit and a basis for whole energy systems analysis. We advocate that having data at the right level of granularity aids in the analysis of localised energy systems, and given the increasingly important role of the electrical system in the delivery of energy, it is logical to propose that geographical units that are meaningful in terms of electrical network boundaries will themselves become increasingly important for the aggregation of underlying data. The concept presented here of using meaningful geographical units for the GB electrical system to aggregate data could translate to different electrical networks around the world that have similar types of service areas supplied by distinct pathways through the electrical networks. Example analyses and use cases have been presented such as the aggregation of publicly available postcode level, annual domestic gas and electricity consumption data to primary substation areas. These analyses resulted in the ability to create helpful maps, for example, to show the mean domestic gas consumption decile of areas served by distinct primary substations, as well as projected future annual demand from the deployment of electric HPs. Discussion of the differences in domestic energy demand across GB's DNO licence areas with weather conditions and socio-economic variables such as affluence and dwelling size would provide even further value to local energy systems modellers. The outputs of this work have been made available to researchers through Zenodo [44] with the intention of providing value to energy system researchers and local energy planners, since greater visibility of electrical infrastructure and related data can inform and contrast the many trade-offs that need to be considered under different decarbonisation options.

Although the analysis was conducted using a smaller number of public datasets which contained intrinsic uncertainties, the analysis has demonstrated the art of the possible; any other data that exists at sufficient granularity of value to the energy system could also be aggregated to the primary substation service areas. Future directions of investigation have been proposed including a consistent method to derive boundaries across GB (a Voronoi approximation using LV substation points and primary to LV parent-child relationships appears to be the most logical for now), the aggregation of further datasets to primary substations (e.g. domestic EPCs to understand where retrofit could have the most impact on creating additional headroom) and ultimately the extension of this analysis to the LV network. The latter task is challenging due to the variability of digitalisation of network geographical data at that voltage level across different DNOs. Nevertheless, it should be seen as a challenge for the sector and academia to overcome by researching scalable, automated or semi-automated methods to create LV service area polygons (which account for logical barriers between infrastructure, substation capacities and the road layout) or through committing resources to digitalising existing information and matching across different datasets such as address data for meters or pre-digitised maps. As more LV monitoring data also becomes available, it would be useful to combine this with the improved LV geospatial data and make it as accessible as possible to researchers and other interested parties, ideally through application programming interfaces to enable near real-time communication and analysis. This would facilitate the creation of digital shadows (or ultimately twins) to enable long-term network planning and sensitivity analysis. Simultaneously, care should be given to protect privacy concerns for small sample sizes of customers; to solve this a redacted version could be made more open and a more detailed version only available to approved parties and the network itself, although it is our opinion that only the monitoring data should be subject to some form of potential suppression or changes to protect privacy in limited cases, rather than the geospatial data. Even internal to the DNOs, the full dataset could be an invaluable tool to aid the active management of LV networks that could help mitigate some of the need for costly reinforcement.

The key recommendation of this research is that meaningful geographic units for energy systems should be used to aggregate datasets so that the data becomes more immediately useful to energy systems analysis. Much of the data available to energy systems analysts at this moment is in units of geography unrelated to the energy system, or rather, there is some relation between the geographical units as most are based on some degree of population awareness. For example, there are data that are aggregated to postcode geographies, but postcodes are useful for the delivery of post, not the investigation of energy systems. Equally data exists for LSOAs but these boundaries are useful for the longitudinal study of population characteristics, not for the investigation of energy systems. Given the growing importance of local energy systems analysis, we therefore advocate that any raw data, where available, should be aggregated to meaningful energy system boundaries as well as postcode and LSOA boundaries. In essence, we are advocating for an additional set of aggregations to be published. This would have to be based on an agreed set of geographical boundaries to allow the aggregation to take place. Postcode boundaries have a robust system of governance, and so do LSOA boundaries. So, in order to support the key recommendation of having data aggregated to meaningful energy system boundaries, there is a further recommendation to set up a robust system of governance around the creation of the boundaries. As part of the governance structure, there could also be a dedicated, public platform; potentially as a subset of an existing service such as the OGP, EDINA Digimap or CDRC, that hosts and maintains a consistent epoch of primary substation polygons. These should be updated periodically as the data improves, i.e., with LV polygons becoming available, or due to changes in network configuration, new connections and upgrades. Engagement with the gas sector to create analogous gas supply areas would be a useful endeavour too. This platform could ultimately then act as a one-stop shop for geospatial energy infrastructure data covering anywhere in GB. Thus, it would prevent the isolation of different datasets across different DNOs (or gas distribution network operators) and be a useful digital resource, well-aligned with GB's ambition to be a leader in both achieving net-zero energy systems and supporting the digital economy.

More globally, the concept of data aggregations to meaningful geographical levels for energy systems analysis (in addition to those for governance such as municipal boundaries) could potentially also be considered by organisations such as the International Energy Agency or online communities such as Open Street Map (who already are working on an Open Infrastructure Map [123]), with the aim to create polygons for those countries where underlying data would permit this to happen.

# 6 Acknowledgements

Funding for this study was provided by the Engineering and Physical Sciences Research Council (EP/W008726/1) under the Gas Net New project and is gratefully acknowledged. Additional support was provided by the Alan Turing Institute under their Science of Cities and Regions Programme to allow for investigation of LV network geography in East Birmingham. The authors would also like to thank the data teams at National Grid Electricity Distribution (formerly Western Power Distribution), UK Power Networks, Scottish and Southern Electricity Networks, Scottish Power Energy Networks, Electricity Northwest Ltd and Northern Power Grid for their support and assistance towards providing datasets, answering queries and giving permissions to place the created resource on Zenodo under open access. Finally, acknowledgement is given to the QGIS Development Team, the geopandas Python library authors and Open Street Map Contributors, whose collective efforts have also made the analysis and presentation of results in this work possible.

# Appendices

## Appendix A

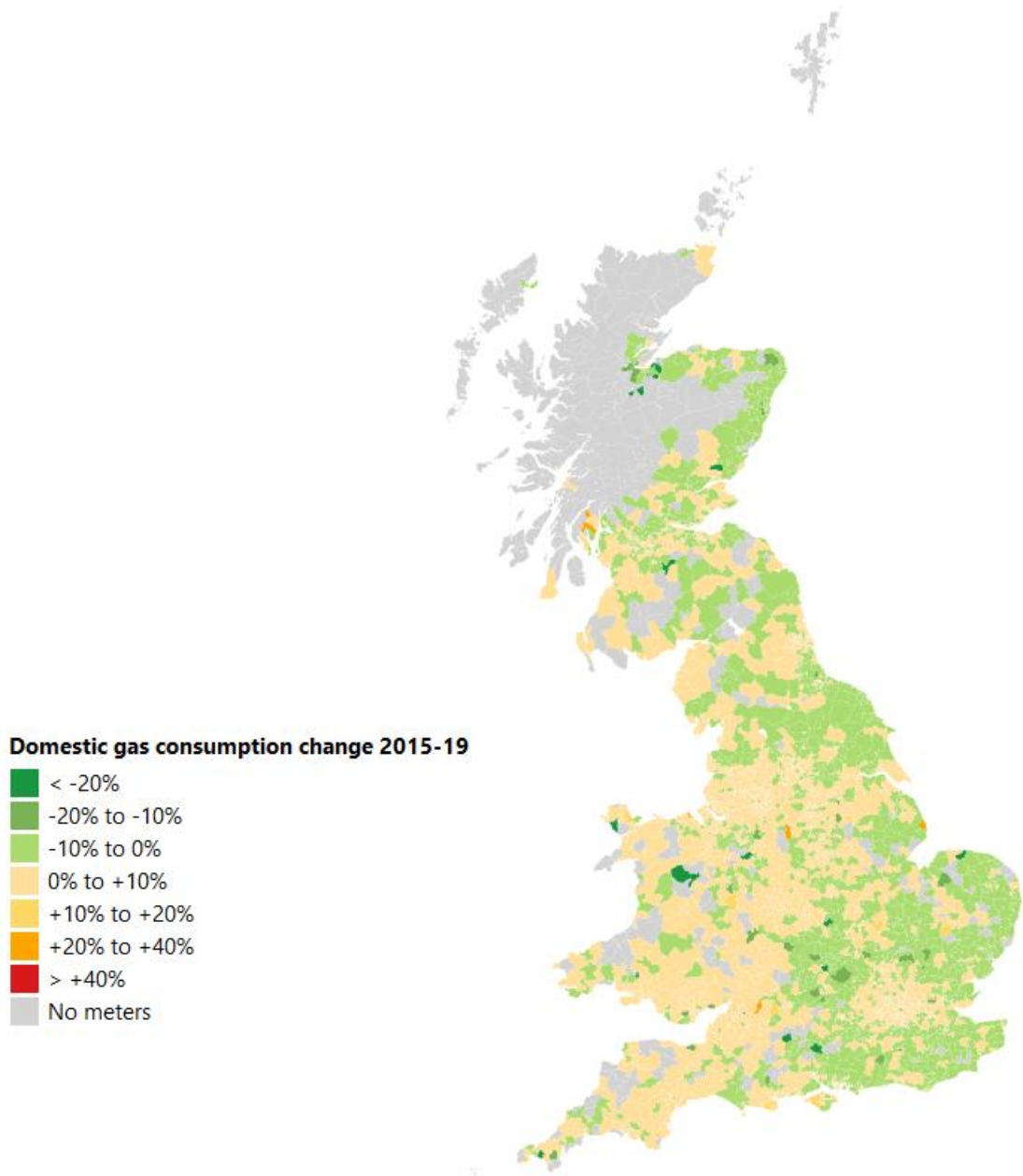

*Figure A.1 - Map to show the change in mean domestic gas consumption between 2015 and 2019 for GB primary substations.*

# Appendix B

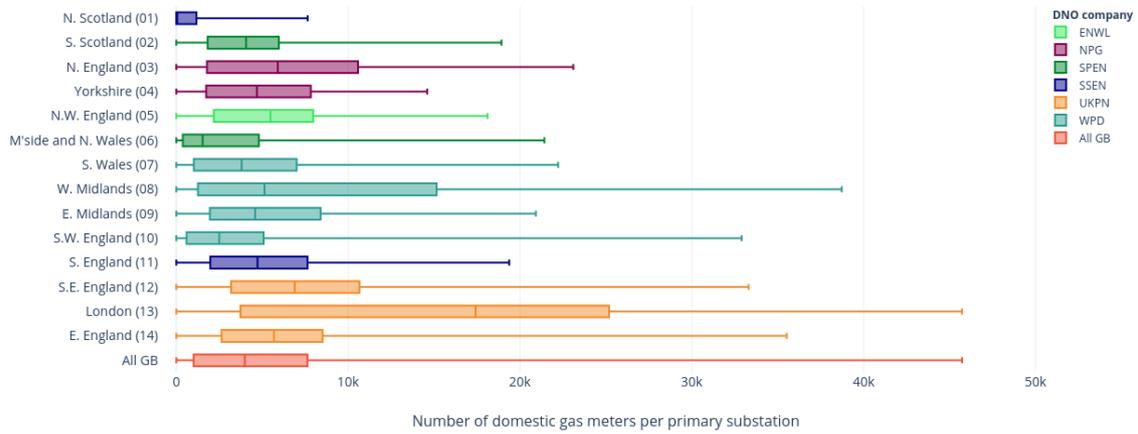

*Figure B.1 - Horizontal box plot to show the distribution of the number of domestic gas meters per primary substation by DNO licence area as well as for all of GB.*

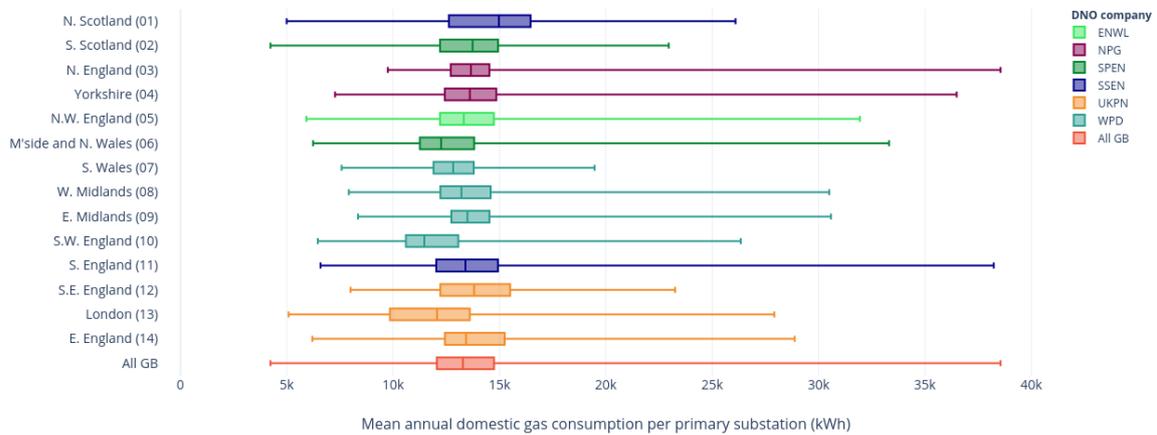

*Figure B.2 - Horizontal box plot to show the distribution of the mean domestic gas consumptions per primary substation by DNO licence area as well as for all of GB.*

# Appendix C

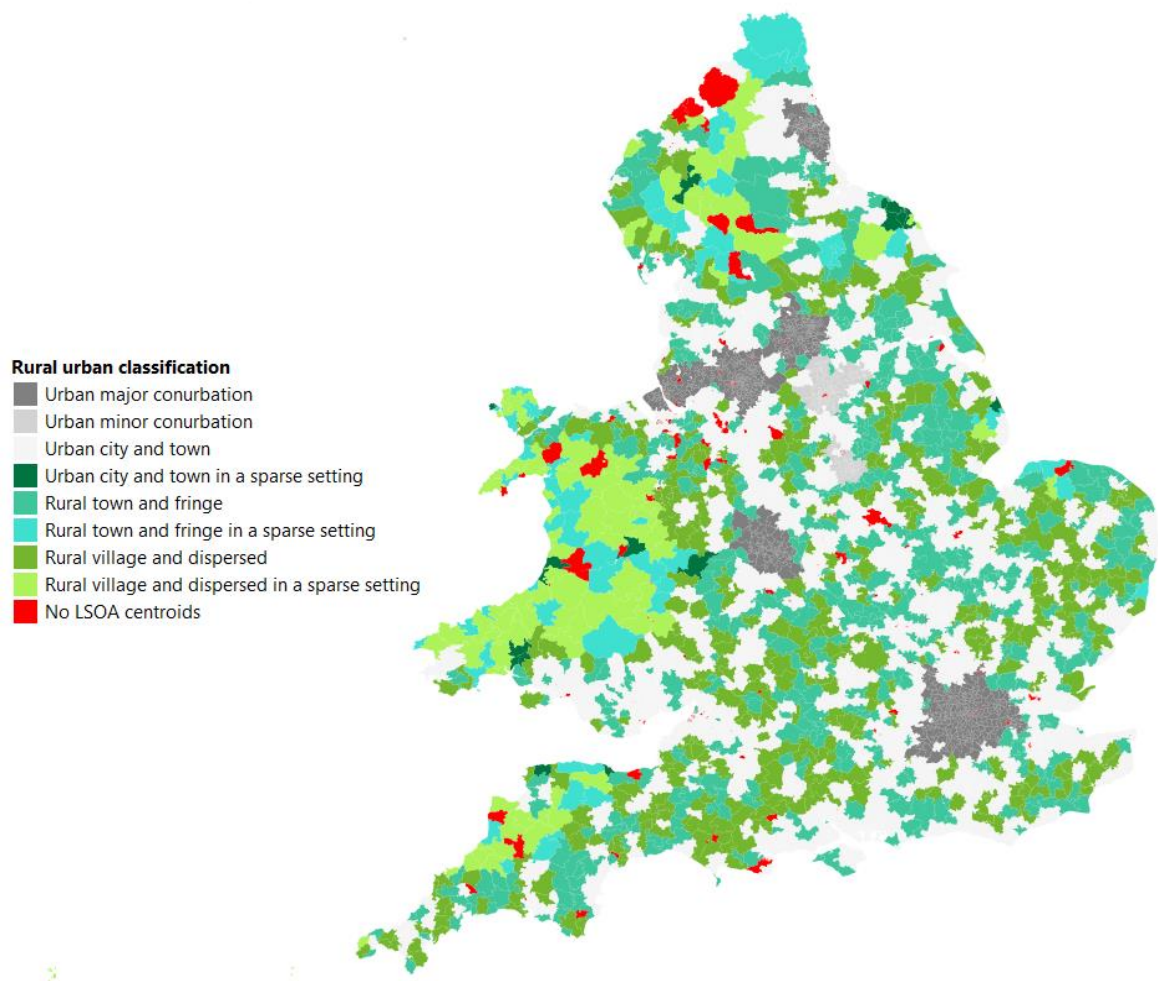

*Figure C.1 - Map to show the RUC of each England and Wales primary substation.*